**At the crossroads of epidemiology and biology: bridging the gap between SARS-CoV-2 viral strain properties and epidemic wave characteristics.**

Florian Poydenot[1], Alice Lebreton[2,3], Jacques Haiech[4*] and Bruno Andreotti[1]

1) Laboratoire de Physique de l'Ecole Normale Supérieure (LPENS), CNRS UMR 8023, Ecole Normale Supérieure, Université PSL, Sorbonne Université, and Université de Paris, 75005 Paris, France

2) Institut de Biologie de l'ENS (IBENS), École Normale Supérieure, CNRS, INSERM, Université PSL, 75005 Paris, France

3) INRAE, Micalis Institute, 78350 Jouy-en-Josas, France

4) CNRS UMR7242 BSC ESBS, 300 Bd Sébastien Brant, CS 10413, 67412 ILLKIRCH cedex

* Corresponding author: haiech@hotmail.fr

Declaration of competing interest: The authors declare that they have no known competing financial interests or personal relationships that could have appeared to influence the work reported in this paper.

Contributions: F.P, J.H and B.A have collectively designed and written the paper, with input from A.L.

ABSTRACT

The COVID-19 pandemic has given rise to numerous articles from different scientific fields (epidemiology, virology, immunology, airflow physics...) without any effort to link these different insights. In this review, we aim to establish relationships between epidemiological data and the characteristics of the virus strain responsible for the epidemic wave concerned. We have carried out this study on the Wuhan, Alpha, Delta and Omicron strains allowing us to illustrate the evolution of the relationships we have highlighted according to these different viral strains.

We addressed the following questions:

1) How can the mean infectious dose (one quantum, by definition in epidemiology) be measured and expressed as an amount of viral RNA molecules (in genome units, GU) or as a number of replicative viral particles (in plaque-forming units, PFU)?

2) How many infectious quanta are exhaled by an infected person per unit of time?

3) How many infectious quanta are exhaled, on average, integrated over the whole contagious period?

4) How do these quantities relate to the epidemic reproduction rate R as measured in epidemiology, and to the viral load, as measured by molecular biological methods?

5) How has the infectious dose evolved with the different strains of SARS-CoV-2?

We make use of state-of-the-art modelling, reviewed and explained in the appendix of the article (Supplemental Information, SI), to answer these questions using data from the literature in both epidemiology and virology. We have considered the modification of these relationships according to the vaccination status of the population.

We hope that this work will allow a better integration of data from different fields (virology, epidemiology, and immunology) to anticipate the evolution of the epidemic in the case of COVID-19, but also in respiratory pathologies induced by a virus or a bacterium transmissible in an airborne manner.

INTRODUCTION

Evidence of aerosol transmission of SARS-CoV-2, the virus responsible for COVID-19 disease, has accumulated over the months (Fennelly, 2020; Lewis, 2020; Morawska and Milton, 2020; Zhang et al., 2020, p. 19) until a consensus was reached, six months after the start of the pandemic. Viral particles, with or without a liquid droplet surrounding them, are dispersed by turbulent air movements. When they are light enough, hydrodynamic fluctuations keep these particles suspended in the air, despite gravity. The mixture of air and particles then constitutes a phase called aerosol. These droplets, which may or may not carry the virus, are produced by atomization in the respiratory tract when an airflow of sufficient velocity causes the fragmentation of a mucus film (Bourouiba, 2021; Johnson et al., 2011; Johnson and Morawska, 2009; Moriarty and Grotberg, 1999). This is the case when large millimeter-sized droplets are emitted by coughing or sneezing, as well as when smaller micron- or submicron-sized droplets are emitted during human exhalation activities (e.g., breathing, speaking or laughing). The water in the droplets then evaporates into the air, concentrating the droplets into virions and mucus proteins, some of which have antiviral properties that help inactivate the virus after a few hours. When airborne viral particles, regardless of their production process, are inhaled, infection can occur (Figure 1). Pathogens responsible for other diseases such as influenza, tuberculosis or measles can also be carried by these small droplets (Blanchard, 1989; Chingin et al., 2018; Du et al., 2020; Zhou et al., 2018).

Airborne particles are the primary route of transmission of SARS-CoV-2 (Cheng et al., 2021; Johansson et al., 2021). The epithelial cells serving as loci of original infection and reservoirs of dissemination are located in the nasal cavity (for strains prior to Omicron), which points to airborne transmission. The heavier, millimeter-sized droplets emitted specifically in the Covid symptomatic phase have a ballistic trajectory that is relatively insensitive to the presence of air and are stopped by all types of face masks and respirators. The reduction of the risk of transmission of the virus outdoors or in well-ventilated closed environments (Gettings, 2021) that is obtained by wearing respirators designed to filter aerosols (Goldberg et al., 2021; Klompas et al., 2021), but also the long-distance transmission in super-spreading events where a single virus carrier infects a large number of people (Endo et al., 2020; Yang et al., 2021), are evidence of the importance of airborne transmission of SARS-CoV-2. Transmission by contact with fomites on which these drops are deposited is probably insignificant (Chen et al., 2021; Goldman, 2020; Lewis, 2021); however, regardless of its actual weight in SARS-CoV-2 transmission, improved hand hygiene remains a recommended habit to prevent transmission of other pathogens — for instance, viruses causing gastroenteritis have a major handheld transmission route associated with a specific locus of infection: the gastrointestinal epithelium (Green et al., 2020; Robilotti et al., 2015). Finally, possible transmission through feces (Nouri-Vaskeh and Alizadeh, 2020) via aerosolization during toilet flushing remains controversial and is probably a minor route, if relevant at all.

Two approaches have been proposed in the scientific literature to characterize the infectivity and transmissibility of viral strains. The epidemiological approach, based on contact tracing and population-scale testing, provides precise quantitative information but has a blind spot: the epidemic propagation depends on social practices, the full complexity of which is difficult to delineate, and which themselves depend on age, education, number and duration of social contacts, vaccination status, etc. It also depends on the degree of immunity in the population. The *ab initio* approach, based on knowledge of virology, immunology, and molecular biology, is complementary; it allows one to characterize viral strains *in vivo* and *in vitro* but suffers from the need of large-scale statistics, and can present important biases of parameterization and calibration. This review article aims to bridge the gap between these two approaches and to review methods combining epidemiological and biological measurements performed on viral strains to deduce their intrinsic characteristics.

Figure 1-A illustrates how the viral mist exhaled by an infected person (index case) can infect non-immune individuals (secondary case) at some distance, and after a time delay. The infection risk increases with the intake viral dose $d$, defined as the amount of inhaled viral particles accumulated over time. $d$ increases with the time of exposure to the virus and with the concentration of infectious viral particles in the inhaled air. The dilution factor between exhaled and inhaled air is controlled at short distance by turbulent dispersion and at long distance by ventilation, which is the process of introducing fresh air into indoor spaces while removing stale air (Poydenot et al., 2022). The dose-response curve expresses the ratio of infected individuals as a function of the intake dose $d$ (Figure 1-A). We hypothesize that a single replicating virion among the numerous particles inhaled can initiate infection. In tissue culture assays, the number of infected cells is proportional to the number of viral cells, which shows that there is no cooperativity between viral particles, *in vitro*(Houng et al., 2004). Then, each inhaled viral particle can be considered as an independent attempt to contaminate the individual. Statistically, more than one is required, as the probability of a single virus successfully overcoming the host immune defenses is low, and since a large fraction of inhaled viral particles, being damaged or defective, are intrinsically unable to infect cells and replicate therein. Infection occurs when a single flawless virion enters a vulnerable location where conditions are permissive to cellular colonization and viral replication. The dose-response therefore follows a cumulative Poisson distribution. By definition, an infectious quantum is the dose inhaled by the individuals of a cohort which leads to the infection of 63% of the cohort. In the field of epidemiology, infectious quanta are used to express a quantity of virus needed to induce contamination or a given symptom (fever, mortality for example).

The dose-response curve has not been directly measured on humans. Infectious challenge trials during which healthy young volunteers are deliberately infected are rare since they are ethically controversial (Adams-Phipps et al., 2022); in the single one led to study COVID-19 (Killingley et al., 2022), a large viral dose (10 TCID50) is introduced via intranasal drops. How can the mean infectious dose (one quantum, by definition) be measured and expressed as an amount of viral RNA molecules (in genome units, GU, Figure 1-B) or as a number of replicative viral particles (in plaque-forming units, PFU, Figure 1-C)? How many infectious quanta are exhaled by an infected person per unit of time? How many infectious quanta are exhaled, on average, integrated over the whole contagious period? How do these quantities relate to the epidemic reproduction rate $R$ as measured in epidemiology, and to the viral load, as measured by molecular biological methods? How has the infectious dose evolved with the different strains of SARS-CoV-2?

In this manuscript, we review concepts and methods providing preliminary answers to these questions. We first describe the mechanism of viral infection and the antiviral response. Then, we detail the viral kinetics and the subsequent time evolution of the viral exhalation flux. The determination of the infectious dose by combining methods from molecular biology and epidemiology is then reviewed. Finally, the evolution of the characteristics of successive viral strains is presented and discussed.

To generate the figures of this review, we have used standard epidemiological modelling. The specific theoretical frame chosen has been published in a previous paper (Poydenot et al., 2022). In the supplementary material section, we provide a detailed model accessible with basic knowledge of physics and mathematics, focusing on the parameter's calibration.

**Mechanism of viral infection**

SARS-CoV-2 is a virus enveloped by a lipid bilayer in which E, M, and S proteins are inserted (Figure 1-B). The lipid membrane originates from the cell in which the virus replicated before being released. The virus contains a copy of the viral genomic RNA protected by a capsid, structured by the assembly of the nucleocapsid protein N. Viral particles measure 80-90 nm in diameter, and are decorated with an average of 48 spike (S) proteins anchored in their envelope. The RNA genome encodes 29 proteins, including the envelope (E, M, and S) and capsid (N) proteins, as well as non-structural proteins required for replication and assembly of the virus within the host cell (Bar-On et al., 2020; Yao et al., 2020). To colonize a cell, the virus interacts via the S protein which is cleaved by a host cell protease (mainly the TMPRSS2 protease for the wild strain, Wuhan-1) with a host cell membrane protein (mainly the ACE2 receptor). This interaction leads to the formation of a virus-ACE2 complex (via the virus spike, a trimer of the S (Yin et al., 2022)) which triggers the internalization of the virus into the cell (Figure 2-A). A series of cellular events leads to the disassembly of the virion and the undressing of the RNA molecule. The released viral RNA is taken over by the host cell ribosomes, which read the information it encodes and produce the viral proteins needed for virus production. New viral particles are then assembled by hijacking the host cell mechanisms, and then released, leading to the colonization of neighboring cells (Snijder et al., 2020). From the nasal cavity or throat, which are probably the first tissues to be infected, the virus, embedded in the mucus secreted by goblet cells of the nasal epithelium, is transported to the trachea, then to the lungs or esophagus, and finally to deeper organs (Figure 2-B and 2-C). The severity and variety of disease symptoms depend on the likelihood of the viral infection overcoming host defenses and reaching multiple sites, as well as on the damage caused to the host by the potent inflammatory and interferon responses launched against the viral assault. In contrast, the spread of the virus depends primarily on its ability to colonize the host airways, and thus the viral load cannot be correlated with symptom severity (Le Borgne et al., 2021). From the nasal cavity, the virus has the ability to travel to the brain and colonize cells of olfactory bulb, which could explain the changes in taste and smell and, in the long term, some of the neurological symptoms associated with long COVID (Fodoulian et al., 2020; Pereira, 2020). More rarely, the virus cpean be found in the blood or lymph, reach different organs in the body and colonize specific cell types in various organs (liver, kidney, heart, prostate, etc.) (Dong et al., 2020) (Figure 2-C).

When the virus is concentrated in the nasal cavity or in the throat, it is spread by a mist of fine droplets of mucus or saliva which can be dispersed by breathing, speaking or singing. A sneeze or cough produces larger droplets containing viral particles. As these droplets are formed, their viral particle content increases linearly with the viral load, in the nasal cavity for mucus droplets, or in the throat for saliva droplets (Buonanno et al., 2020b). An organism can be infected if enough viral particles interact with cells expressing both the TMPRSS2 protease and the ACE2 receptor and if the virus is able to hijack cellular mechanisms to produce and disseminate new virions.

For the Omicron viral strain, the TMPRSS2 protease appears to be less essential, in favor of an alternative pathway of entry into epithelial cells via the endosomal route (Peacock et al., 2022b). At each cycle, the virus must enter a novel host cell, replicate its RNA molecule, produce the proteins necessary for its self-assembly and then be released as virions.

Omics databases such as the Human Cell Atlas (https://www.humancellatlas.org/) provide insight into the tissues and cell types that express the TMPRSS2 protease and ACE2 receptor in different tissues of the human body. It is commonly assumed that these cells are the primary target cells of SARS-CoV-2 (Hoffmann et al., 2020). The nasal epithelium is most often the first infected tissue, which is amongst the strongest proofs in favor of a transmission of SARS-CoV-2 primarily by inhalation of virus-laden aerosols. Accordingly, there exists a cluster of nasal cells in the ACE2/TMPRSS2 expression map (Singh et al., 2020; Sungnak et al., 2020). However, due to its lower dependence on TMPRSS2, the Omicron viral strain illustrates the changes of cellular tropism that can happen as a result of evolution of the virus (Gupta, 2022).

**Antiviral response**

The viral load of an infected person is the result of viral replication and of the antiviral response. The arrival of the virus in the nasal cavity or in the throat induces an antiviral response, due to the recognition of viral components considered as danger signals by relevant cellular receptors (Silva-Lagos et al., 2021). This interaction triggers a non-specific antiviral defensive response, which relies to a large extent on the production of a class of molecules called interferons. By binding to their receptors on target cells, interferons induce the expression of a wide range of interferon-stimulated genes (ISGs) that, through various mechanisms, help limit viral replication (Gallo et al., 2021).

Interferons are produced by infected cells but also by sentinel immune cells such as macrophages and dendritic cells. After their secretion, they diffuse and bind to their receptors on surrounding cells without discriminating whether they are infected or not, stopping cellular functions. Since the concentration is higher around the site of infection, diffusion leads to an efficient stochastic control of infected cells. If the interferon response is initiated early, in a localized and circumscribed manner, viral spread through the tissue can be controlled (Katze et al., 2002; Perry et al., 2005). However, a too strong interferon response is not only antiviral but also destructive to host cells. Moreover, the outcome of the virus-host crosstalk is further complicated by the fact that some viral proteins counteract host interferon responses and, conversely, host interferon responses can sometimes amplify viral infectivity (Ziegler et al., 2020). Others mechanisms may also modulate the infectivity of specific strains, for instance the modulation of splicing of ACE2 in response to interferon responses (Blume et al., 2021).Therefore, the nasal cavity is a site of ongoing conflict between SARS-CoV-2 replication and the body's efforts to inhibit the virus. The efficiency of the viral replication and secretion processes, as well as the body's interferon response, can vary across time and space within the nasal cavity. These above described and antagonists processes can influence the ability of the virus to establish an infection in the body.

Because viral transmission depends directly on viral loads, the kinetics of the replication and secretion of the virus characterizes the capacity of an individual to transmit the infection. Moreover, the kinetics of the interferon response correlates with an individual's susceptibility to infection (Hadjadj et al., 2020). Although it is often forgotten, mucus also modulates significantly that kinetics in the nasal cavity and participates in the immune response. Mucus is composed mainly of water (95%), lipids and glycoproteins (mucins) (Bansil and Turner, 2018; Fahy and Dickey, 2010). It is secreted by specialized cells named goblet cells which are also among the cells infected by the virus. Mucus forms two layers at the surface of the nasal epithelium: a gel constituted by a network of polymerized mucins anchored to the membrane of epithelial cells, and a solution that moves under the action of the cilia of the epithelial cells. The viscoelastic properties of mucus depend on the type of mucins (Moniaux et al., 2001) and on the physicochemical environment (humidity, calcium concentration and pH). Inflammation modifies the expression and the glycosylation/fucosylation of MUC5B and MUC5AC (Amini et al., 2019; Chatterjee et al., 2020), the major airway gel-forming mucins.

Mucosal immunity results from a synergy between the mucus and the immune system (innate immunity, secreted IgA and IgA for the acquired immunity, and cellular immunity), as the network created by polymerized mucins traps particles down to a few hundred nanometers in size (which is the size of viruses). Mucus also contains enzymes that can neutralize viruses and bacteria in a non-specific manner. It should be noted that nasal mucus and saliva, which is also a specific mucus, have different compositions and physicochemical characteristics. Lastly, the nasal microbiota, a community of bacteria that live within the mucus, also plays a role in defense against infection by viruses or pathogenic bacteria (Kolhe et al., 2021).

**Viral load**

The viral load in the oropharyngeal cavities, which determines the exhaled particle flux, is highly time-dependent (Figure 3-A). It can be measured in different ways. The concentration of viral genomes in a sample can be measured using RT-qPCR (Reverse transcription followed by quantitative Polymerase Chain Reaction), using as targets one or more nucleic acid sequences in the viral RNA genome (Figure 3-B). The sample is collected with a swab, and then its RNA contents are extracted in a given volume of an RNA extracting and stabilizing solution. This volume differs depending on the test used. The RT-qPCR result is expressed in threshold polymerase reaction cycles (Ct) which is transformed into a number of copies of viral RNA molecules (genome units, GU), after calibration using a solution of viral RNA molecules of known concentration. The concentration of viral genomes is thus expressed in viral RNA molecules per unit of volume of extraction solution (GU/mL in practice). Considering that the swab collects at most 100 microliters of nasal extract and the volume of extracting solution varies between 1 and 3 mL depending on the kit used, the concentration of viral RNA is probably an order of magnitude higher in the oropharyngeal cavity than the concentration in the extraction solution. Alternatively, the ability of the virus to infect a cell monolayer and cause cell lysis can be measured. Using serial dilutions of a virus-containing sample, the number of lysis plaques (Plate Forming Unit, PFU) produced within a well monolayer for a given volume of sample is measured (Figure 3-C). The concentration of replicative virus expressed in PFU/mL depends on the type of cells used, as susceptibility to infection varies depending on cell types.

For a given sample and a given cell type, the number of cell lysis plaques obtained is proportional to the concentration of viral genomes in solution. This experimental linearity indicates that cell infection events can be considered as independent from each other. The number of lysis plaques induced on the average by one viral genome unit is the viral infectivity and is expressed in PFU/GU (Dabisch et al., 2021; Wang et al., 2021). As this quantity reflects the ability of the virus to enter cells and replicate, it primarily depends on the viral strain considered and on the type of cells. In practice, the viral infectivity is a highly variable quantity as it reflects viral integrity for a given sample. It therefore depends on the time and site of sampling, as well as on the protocol used for sample collection and preservation. A viral infectivity equal to one would imply that each virion is able to colonize a cell, replicate, be released to colonize neighboring cells and lyse the colonized cells. This value is never reached as it implies a perfect replication and assembly of all virus particles, the absence of damaged or inactivated viruses, as well as perfect sample preservation. In other words, low values of the viral infectivity indicate viral solutions that contain high loads of mis-assembled, non-functional or degraded viral particles. Part of this degradation may be due to the antiviral response or the immune response of the body. Conversely, a null viral infectivity means that no virion can form lysis plaques on a cell monolayer. For SARS-CoV-1, the viral infectivity was between $6 \times 10^{-4}$ and $8 \times 10^{-4}$ PFU/GU on VERO cells (Houng et al., 2004) : one viral particle over 1200 to 1600 was able to form a cell lysis plaque. For the Wuhan-1 strain of SARS-CoV-2, the literature provides different measurements of infectivity for high preservation quality samples (Bao et al., 2020; Yu et al., 2020), which lead to an estimate of $7 \times 10^{-4}$ PFU/GU (1 cell lysis plaque for 1400 virions) for the initial, asymptomatic phase of the pathology. As the disease progresses, the fraction of replicating virions decreases and the infectivity decreases.

The viral kinetics has standardly been described in the literature by an exponential growth of the viral load (replication and shedding) followed by an exponential decay (immunity response). The SARS-CoV-2 "human challenge" trial (Killingley et al., 2022) has provided unprecedented time-resolved data showing a more rounded viral load curve (Figure 3-A). The viral strain used in this study was close to the wild-type strain, Wuhan-1. In the nose, the maximum viral load, expressed in genome units, was reached after $T_m$=7.0 days, around the onset of symptoms. This time lapse provides an estimate of the incubation period. This period was slightly smaller, around $T_m$=5.9 days, when considering the maximum replicable viral load, measured in PFU. This difference may be ascribed to the effect of the antiviral responses on infectivity. As an alternative hypothesis, we cannot exclude the possibility of incorrect assembly of the virus by the colonized cell leading to a decrease of infectivity (in PFU/GU), although a potential inhibition of the virus assembly could also be a consequence of antiviral responses. The infectious period $T$, defined as the average time between infections of the index and secondary cases, is estimated to be 7.2 days (Figure 3-A). The exponential growth rate just after infection is approximately 15.0 days$^{-1}$ (growth time 8.5 hours). The exponential decay rate long after this maximum is 3.3 days$^{-1}$ (decay time 39 hours). The infectivity is around $10^{-5}$ PFU/GU at maximum, which is two orders of magnitude weaker than the value found when using virus replicated on cultured cells.

The viral emission rate, defined as the quantity of viral particles exhaled per unit of time, has been measured in several pioneering studies. In the study by Ma et al. (Ma et al., 2020), patients were asked to exhale into a cooled hydrophobic film through a long straw to collect a sample of exhaled breath condensate. The measured concentration of SARS-CoV-2 viral particles was between $10^5$ and $2 \times 10^7$ GU/m$^3$. The authors noted that the exhalation rate was correlated with the viral load in the nose and in the throat but not in the lungs. In the study by

Malik et al. (2021), the mean viral load per swab was $7.8 \times 10^6$ GU whereas exhaled breath samples displayed $2.47 \times 10^3$ GU per 20 times exhaling, which corresponds to $2.5 \times 10^5$ GU/m$^3$. In the scientific literature, it is standardly considered that the viral emission rate is proportional to the viral load. This is reasonable if the infected epithelium surface is reasonably constant and if there is no viral enrichment at the interface between muco-salivary fluid and air. In the following sections, we will adopt this hypothesis and use the multiplicative factor 0.4 between the viral load, in GU/mL, and the viral emission rate, in GU/ day found in Malik et al. (2021).

The total viral emission of an infected person is the quantity of viral particles exhaled during its infection. Considering the viral kinetic displayed in Figure 3-A, the total viral emission, obtained by integrating over time the viral emission rate, would be around $9 \times 10^7$ GU, or $6 \times 10^4$ PFU, within a half-decade uncertainty.

**Molecular *vs* epidemiologic determination of the infectious dose**

As mentioned in the introduction, the determination of the infectious dose *a priori* requires measuring a dose-response relationship. Importantly, infection can be defined from different observables (amount of virus in the nasal cavity, presence of specific symptoms — like rhinitis, pneumonia or acute respiratory distress syndrome — or mortality). The infectious dose therefore depends on the viral strain, on the capacity of the upper airway mucosal immune system and of the systemic immune system for the lung or other organs to neutralize the virus, and on the type of observables used to monitor infection. The inhaled dose only dictates the primary infection in the upper airways. The severe acute respiratory syndrome in the lungs typically results from a secondary infection, after the virus has colonized the nasal cavity (Wölfel et al., 2020). As a consequence, the relevant viral dose is the one that is transferred from upper to lower airways. The viral dose in the *inoculum* is therefore not directly related to disease severity, as it is negligible compared to the production of virus by colonized host cells in the upper airways.

The dose response relationship allows expressing the mean infectious dose (one quantum, by definition) as an amount of viral RNA molecules or as a number of replicative viral particles. However, it can only be measured on animal models. No such study exists for SARS-CoV-2, to the best of our knowledge, but it has been measured on a non-human primate model for SARS-CoV-1 (Watanabe et al., 2010). The ID50 (median infectious dose) describes the amount of replicable virus (expressed in PFU) needed to infect 50% of the population (Figure 1-A). Note that the same quantity, when measured on a population of cells in tissue culture assays, is called TCID50 (median tissue culture infectious dose). For SARS-CoV-1, the reported measurement is ID50=280 TCID50. We recall that an infectious quantum is the dose inhaled by the individuals of a cohort that leads to the infection of 63% of the cohort, and that it is used in epidemiology as a unit. This convention leads to a multiplicative factor $1/\log(2)=1.44$ between the epidemic quantum and the ID50 quantity. The mean infectious dose therefore relates an epidemiological quantity to a characteristic quantity of virus, defined using molecular biology experiments (Figure 1): the reported measurement for SARS-CoV-1 leads to 400 PFU/quantum. Using these estimates, the infectious dose for the raw Wuhan-1 strain on humans is in the range of $5.6 \times 10^5$ GU, within a factor 2. Considering the viral kinetic displayed in Figure 3-A, the total viral emission $\hbar$ would be around 150 quanta, within a factor 6. This corresponds to a typical viral emission rate of 1 quantum/hour, and a maximal viral emission rate of 2 quanta/hour.

From the epidemiological point of view, reference points are provided by closed micro-societies inside which the virus propagates. They provide estimates of the total viral emission $ℏ$ expressed in infectious quanta. In other words, the exhaled dose of an infected person as the potential to infect $ℏ$ other people, on the average. However, thanks to ventilation, only a small fraction of this exhaled dose is actually inhaled. The exponentially growing epidemics onboard the ship Diamond Princess (Almilaji and Thomas, 2020) and onboard the French aircraft carrier Charles de Gaulle are the most important events for the Wuhan-1 strain (Figure 3-A). Both events were characterized by a low rate of replacement of stale air by fresh air, and by an air conditioning system lacking HEPA filters to remove pathogens. Using the viral load curve of Figure 3, the growth rate can be converted into an epidemic reproduction rate (R=4.8 and R=3.2, respectively). Then, using a model dilution factor between exhaled air and inhaled air (Poydenot et al., 2022), one deduces the total viral emission $ℏ$=490 quanta (retired people) and $ℏ$=460 quanta (young adults), respectively. This corresponds to a typical viral emission rate of 3 quanta/hour, and a maximal viral emission rate of 6.3 quanta/hour. Schools constitute the best documented social sub-system (Bazant and Bush, 2021; Vouriot et al., 2021), although not isolated from the rest of society as ships were. Figure S4 shows the epidemic evolution for secondary school pupils, between lockdown and holidays, in UK, together with the typical $CO_2$ concentration measurement (Poydenot et al., 2022) that gives a total viral emission $ℏ$ = 270 quanta for pupils from age 10 to age 14. It corresponds to a typical viral emission rate of 1.6 quanta/hour and a maximal emission rate in the range of 3.3 quanta/hour.

In conclusion, although calibrations are lacking large-scale statistics, the molecular and epidemiologic determinations of the mean infectious dose are consistent with each other, within error bars. It is important to note that the infectious dose is approximately $6 \times 10^5$ GU for the wild type strain Wuhan-1 and not between 10 and 100 as mentioned in a series of recent articles (Bazant and Bush, 2021; Buonanno et al., 2020a; Lelieveld et al., 2020; Pöhlker et al., 2021; Vouriot et al., 2021; Vuorinen et al., 2020). A large part of the results on the airborne transmission risk, based on the volume of exhaled drops, is quantitatively flawed by the omission of the infectivity ratio between plaque forming units (PFU) and genomic units (GU). The estimated typical emission rate is between 15 and 30 quanta/hour in Bazant and Bush (2021) and Miller et al. (2021) found 1,000 quanta/hour for singing, to be compared to 1-3 quanta/hour estimated here. Accordingly, the total viral emission $ℏ$, which ranges between 450 and 500 quanta on average for adults, is almost 10-fold smaller than previous estimates.

**Evolution of viral strain characteristics**

In the previous sections, we have discussed at length the absolute characteristics of the wild strain, Wuhan-1. Comparing the characteristics of the variants presents different difficulties. On the one hand, relative measurements are much more precise; on the other hand, the immunity induced by infection or by vaccination induces a strong heterogeneity in the population. The evolution of the parameters presented here for the wild type strain, Wuhan-1, and for variants Alpha (B1.1.7), Delta (B.1.617.2), Omicron BA.1 and Omicron BA.2, is based on relative measurements, the wild type strain serving as a reference. We have selected the variants that have led to an epidemic wave and we have retained the date of the first case in France to display the results. Although this choice is somehow arbitrary, it is justified by the fact that most measurements are performed using French epidemic data.

Figure 4-D shows that the incubation period $T_m$ and the infectious period $T$, measured *in vivo* using the viral load kinetics, are almost equal. They are slightly decreased for the variants Alpha and Delta, which leads to a higher growth rate and therefore to an evolutionary advantage. By contrast, the advantage of Omicron is not due to a change of incubation and infectious period. Figure 4-D compares these two characteristic times, $T_m$ and $T$, to the replication time, measured *in vitro* using human nasal epithelial cells (hNECs). To get comparable orders of magnitude, we define the latter as the time needed *in vitro* to multiply the number of viral particles by 1 billion. Interestingly, the replication time of the virus in a human lung cell line that expresses abundant ACE2 and TMPRSS2 (Calu-3) has strongly decreased for the Delta variant, in direct relation with a higher risk of pneumonia. Omicron is milder than Delta (but more severe than the wild-type strain) for lung symptoms, which is coherent with its increased replication time in Calu-3 cells.

In Figure 5, we propose a meta-analysis of various characteristics for the successive variants of interest. It combines epidemiologic and biological data, following standard methods in both domains. We emphasize that this meta-analysis is deduced from the definitions and measurement methods defined previously in a straightforward and robust way.

Figure 5-A shows the evolution of the inverse of infectivity, measured as the average number of viral particles (GU) needed to induce one lysis plaque on a generic Vero cell culture (PFU) (data from Bao et al., 2020; Despres et al., 2022; Ghezzi et al., 2020; Houng et al., 2004; Killingley et al., 2022; Paton et al., 2021; Peacock et al., 2022; Peña-Hernández et al., 2022; Puhach et al., 2022); measurements performed on hNECs and Calu-3 cells would be more relevant but are lacking. The Delta variant is more infectious (240 GU/PFU) than the Alpha variant (1,000 GU/PFU), itself being more infectious than the wild-type strain (1,400 GU/PFU). Omicron is slightly less infectious than Delta (300 GU/PFU). This confirms the evolution of the replication time. The evolutionary advantage of the Delta variant is due to its higher binding affinity with ACE2, which increases the probability of cell entry and replication.

Figure 5-B shows the total viral emission $\hbar$ (in quanta), deduced from epidemiologic measurements. The reference value of $\hbar$ is deduced from the infections onboard the Diamond Princess and Charles de Gaulle boats (Figure 3-A). $\hbar$ is the key biological characteristic that determines the epidemic growth rate, for given social practices. The ratio of $\hbar$ for two successive strains is deduced from the period of time during which these two strains coexist, as the ratio of the epidemic reproduction rates $R$. The two values of $R$ are themselves deduced using the Euler-Lotka equation from the epidemic growth rates. Figure 5-B shows that $\hbar$ has increased from one variant to the next. For Alpha and Delta, this increased epidemic growth rate is probably a direct consequence of the evolutionary increase of binding affinity for the ACE2 receptor. For Omicron, a reasonable hypothesis is that the increased epidemic growth rate results from a displacement of the first entry point from the nasal cavity to the throat. A possibility would be that lower density of lymphid follicles in the throat than in the nasopharynx renders this mucosa less potent in launching an effective immune response against infection (Ogra, 2000). The total viral emission $\hbar$ (in quanta) is also displayed when both the index and secondary cases are fully vaccinated. For this, we used data of the relative transmissibility and susceptibility deduced from household transmission. Although the vaccines were optimized to induce circulating antibodies and systemic T and B cell responses to prevent viral diseases, they were very effective to prevent transmission for the Alpha and Delta strain. On these variants, the vaccines were reducing both the transmissibility, which is by definition the replicable viral emission rate, and the susceptibility of contact cases to be infected. However, this prevention of viral spreading by the vaccine has almost disappeared with Omicron. Incidentally, by maintaining a broad viral pool in circulation, increased immune escape is amongst the reasons for Omicron increased transmissibility (Paz et al., 2022).

Figure 5-C shows the evolution of the infectious dose (1 quantum) in the upper airways, in genome units (GU). This measurement combines the number $ℏ$ of infectious quanta exhaled during the infectious period (Figure 5-B) and the evolution of the maximum viral load. The ratio between the number of genome units in a swab and in one breath is around $1.6 \, 10^5$ (Adenaiye et al., 2021; Leung et al., 2020; Ma et al., 2021; Malik et al., 2021; Ryan et al., 2021). The result is multiplied by the mean number of breaths per day ($2 \times 10^4$) and by the integral over time of the viral load in the upper airways to obtain the number of genome units (GU) exhaled on the average during the whole infectious period. The number of genome units in the infectious dose is obtained by dividing the number of genome units (GU) exhaled in total by the number of quanta exhaled in total, $ℏ$. Figure 5-C shows that the reason for the increase of the integral viral exhalation $ℏ$ is not primarily a larger viral load but a strong decrease of the number of virions statistically needed to induce the infection.

Figure 5-D shows the evolution of the infectious dose (1 quantum) in the lower airways, measured in genome units (GU), for the successive variants of interest. The mean dose required to infect the lower airway is measured as the ratio of the number of genome units (GU) transferred by inhalation from the upper airways to the lungs, divided by the probability to develop a lung pathology. In first approximation, inhalation moves the same quantity of viral particles towards the lungs as exhalation does towards the outside. The probability that the lungs get infected by SARS-CoV-2 when the upper airways are, is approximated by the hospitalization hazard ratio. The infectious dose is much larger for the lungs than for the upper airways due to the fact that viral particles are not diluted when transported to the lower airways (Heyder, 2004; Hinds, 1999). The rather low probability of infection shows that the immune system is more effective in the lungs than in the nose or throat.

Figures 5-E and S8 shows the maximum likelihood phylogenies inferred from spike nucleotide sequences for major SARS-CoV-2 lineages. SARS-CoV-2 does not currently show the unbalanced, unidirectional phylogenetic tree that is a hallmark of immune escape for viruses under a strong immune selection pressure, and where each new variant emerges from the last dominant variant (Volz et al., 2013). Indeed, the Omicron strain is not a mutation of the Delta strain, nor was the Delta strain a mutation of the Alpha strain, but each one of them emerged from very different branches of the phylogenetic tree. So far, each new variant of concern has arisen from very different branches of the phylogenetic tree by novel mutations that have remained undetected over long periods of time, resulting in a relatively balanced tree. By contrast, endemic viruses such as influenza or seasonal coronaviruses can be recognized by the shape of their unbalanced, unidirectional phylogenetic tree (or ladder-like evolutionary tree, when represented as a function of time): they circulate and mutate; in parallel, immunity develops against them, leading to the gradual extinction of the ancestral branches. Some of the mutations lead to new variants that escape immunity; their branches expand until immunity develops against these new variants, and so on. At this endemic stage, each new variant thus derives by mutation from one of the last hegemonic strains, usually by a limited evolutionary jump that allows it to escape immunity, which is partially predictable(Carabelli et al., 2023). Meanwhile, at the present stage of SARS-CoV-2 evolution, new viral strains do not present such a phylogenetic pattern: they may emerge

from all branches as well as from the root of the tree, after having spread silently for a time. It is then impossible to predict how close the next hegemonic variant will be to the previous one.

**Discussion**

In this article, we have first reviewed a series of epidemiological and molecular biology methods that can be combined to characterize the airborne transmission of respiratory viruses:

(i) replication kinetics of viral strains in tissue culture assays, using RT-qPCR and cell lysis plaques measurements, provides access to the viral replication time in the absence of immunity response and to the viral infectivity

(ii) the dose response curve, determined using model animals close enough to humans, provides the expression of the epidemic quantum in GU and PFU

(iii) the viral exhalation rate can be measured as a function of time $t$ after infection, directly, by collecting virus in a mask, or via viral load curves. This measurement gives access to the total viral exhalation in GU and in PFU, to the infection time $T$ and to the rescaled viral transmissibility $\psi(t)$

(iv) the epidemic reproduction rate $R$ can be deduced from the epidemic growth rate $\sigma$, using the rescaled viral transmissibility $\psi(t)$. This measurement gives access to the total viral exhalation in quanta, if the epidemic growth rate $\sigma$ is measured on an isolated micro-society

This toolbox can be used to characterize viral strain and calibrate quantitatively the models for airborne transmission risk.

Some limitations to the present conclusions of this study should be highlighted. The severity of the symptoms and the transmissibility depend on the personal status (age, weight, immunodeficiency conditions, comorbidity factors, etc.). Epidemic databases annotated with such information do not exist as open data for obvious ethical reasons: it would be necessary to chain different databases. Such an investigation would be extremely interesting but is outside the scope of this review.

A better molecular determination of infectious quanta requires measuring the number of viral particles per unit of time (or per unit of volume of exhaled air) during expiratory human activities such as breathing or speaking. This would require the design and calibration of face masks that allow patients to breathe normally and collect all viral particles in a filter. We have used here epidemiological data to estimate the infectious quantum, expressed in viral RNA (GU), for the successive SARS-CoV-2 variants. The systematic use of respiratory aerosol samplers (Li et al., 2021) is essential to characterize quantitatively SARS-CoV-2 strains, which is necessary for risk assessment and subsequent risk reduction policies. Similarly, standard quantitative molecular biology techniques, such as plaque assays and RT-qPCR could be used directly to measure face mask efficacy.

Second, it would be useful to measure *systematically* the replication kinetics of SARS-CoV-2 strains in tissue culture assays with large enough statistics and controls to infer replication properties in the absence of immune response. Although viral multiplicative curves are regularly measured and published, the necessary scientific coordination is missing as well as systematic comparison with epidemiologic data, for which the immune response is present.

Third, another limitation of this study is the lack of knowledge about the generation of virus-containing aerosols in the upper respiratory tract, particularly in the nasal cavity. We have shown that indirectly determining the exhalation rate of viral particles using the aerosol droplet emission rate underestimates the result by three orders of magnitude (Duan et al., 2021).

Fourth, several issues regarding evolution during desiccation of mucus droplets carrying virions remain open:

- The evolution of the physicochemical characteristics of mucus as a function of time, temperature, humidity, pH, ionic concentration (especially calcium, which condenses mucin polymers at high concentrations), and pathology (Ma et al., 2018).

- The role of mucus in the formation of droplets and their contaminating character: concentration of the virus, size of the droplets, mixing of the mucus of two different individuals—mucus of the transmitter and mucus of the receiver (Edwards et al., 2021).

- The respective contributions of mucus and interferon responses to the clearance of SARS-CoV-2 (Persson, 2021).

Finally, the mechanisms of SARS-CoV-2 inactivation remain poorly understood. In order to assess the effectiveness of alternative risk reduction techniques, it is necessary to know how environmental conditions (temperature, humidity, chemical concentrations, ultraviolet irradiation) affect the viability of SARS-CoV-2 (Duan et al., 2003; Fears et al., 2020; Lednicky et al., 2020).

The evolution of the virus takes place under a double selection pressure, an increase in transmissibility and an escape from neutralizing antibodies. The increase in transmissibility is due to an optimization of the virus ability to replicate in the epithelial cells of the throat and nose, which release more and more replicable virions. This optimization occurs independently of the symptoms that the different mutants may cause in the contaminated organisms. The existence of neutralizing antibodies affects the quality of the exhaled virions. Thus, immune escape of new variants participates in a temporary increase of transmissibility by increasing the quantity of replicable exhaled viruses.

The severity of symptoms induced by a given variant is not correlated with the transmissibility and therefore, does not seem to be important in the evolution of the virus (Alizon and Sofonea, 2021). Thus, new variants appear randomly, which may either induce more severe, or milder, symptoms. However, the speed and diversity of virus evolution is correlated with its circulation flow and its ability to remain present in an organism for long periods. Thus, the ability of new variants to infect animals or persist in immunocompromised individuals accelerates virus circulation and the appearance of new variants with enhanced transmissibility. The greater the diversity of these new variants, the greater the possibility of variants inducing severe symptoms, even in younger individuals. Following a precautionary principle, it is important to decrease as much as possible the circulation of the virus by improving the air quality in closed areas and to monitor the circulation of the virus and the appearance of new variants in a given spatial territory. This requires appropriate means allowing a timely monitoring of viral circulation in the environment and not only at the level of individuals (Rios et al., 2021).

**Figure captions**

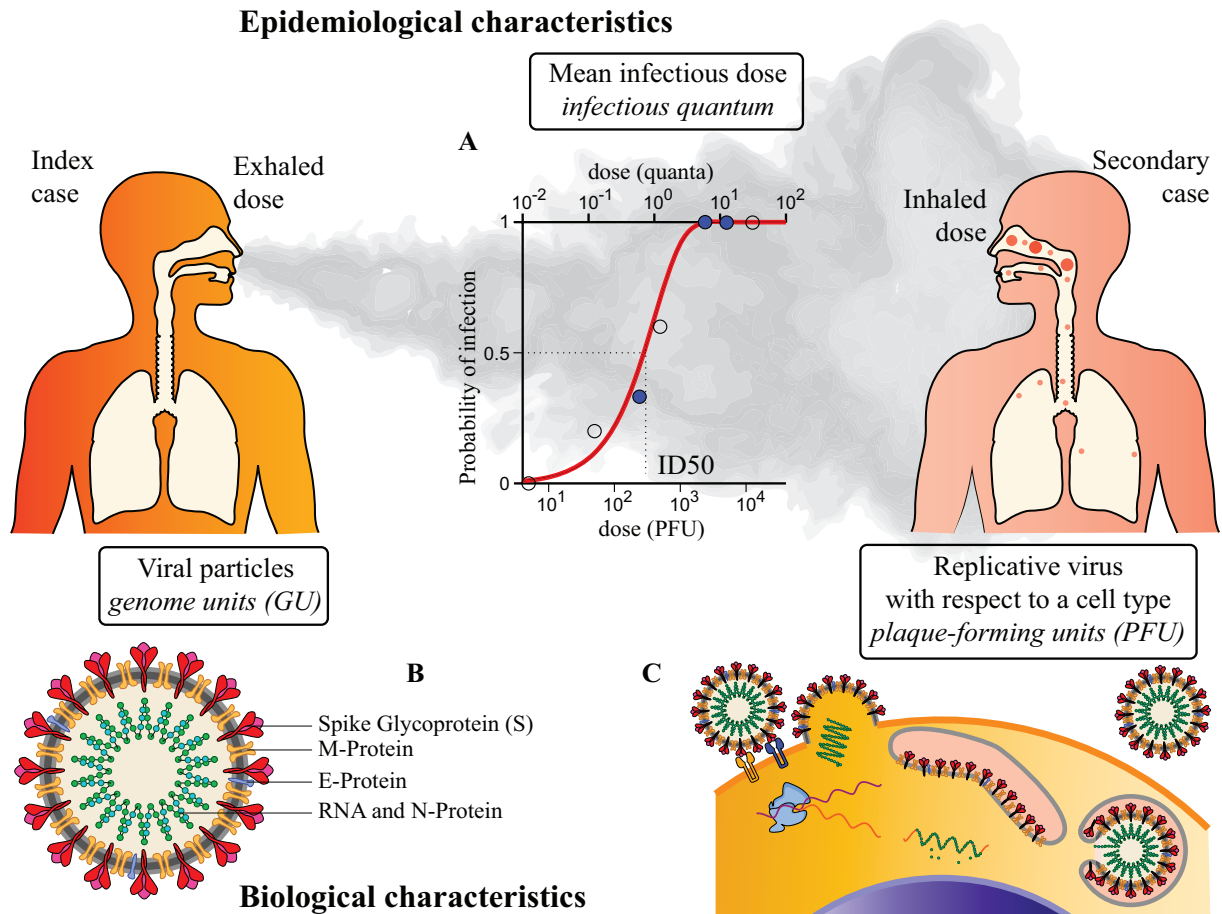

Figure 1: How to relate epidemiologic characteristics to measurements performed in molecular biology? (A) The dose-response curve relates the probability of infection to the amount of inhaled viral particles accumulated over time, called the intake dose. The mean infectious dose is defined as ID50 for animals (50 % probability), as TCID50 for cells (50 % probability) and as the infectious quantum for humans (dose-response curve approximated by 1-exp(-d), where *d* is the dose expressed in quanta; red curve). (B) The mean infectious dose can be expressed as a quantity of viral genome copies, expressed in genome units (GU). (C) Alternatively, it can be expressed as a quantity of viral particles able to replicate on a certain type of cell, expressed in plate forming units (PFU).

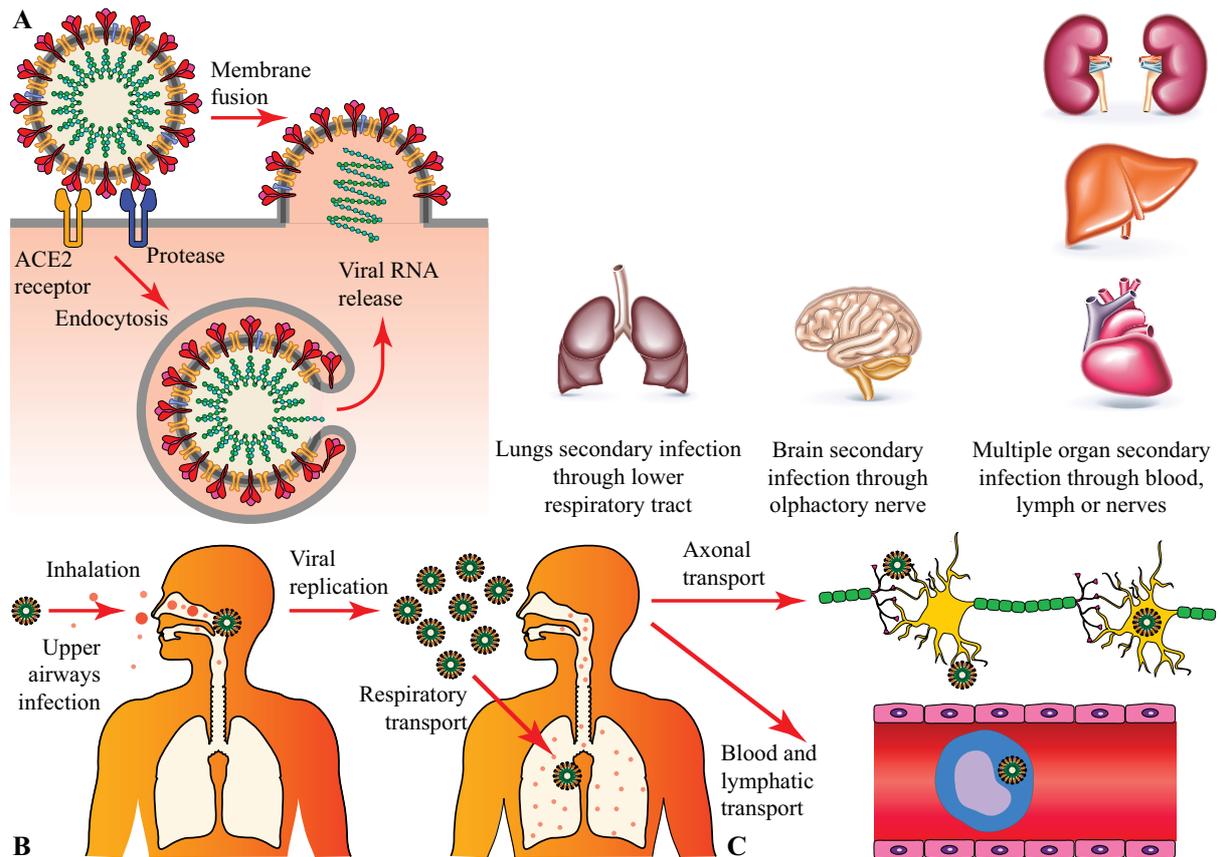

Figure 2: Mechanism of viral infection. (A) Pathway used by SARS coronavirus 2 (SARS-CoV-2) to enter and infect the cell by intermolecular interactions between the spike protein of the virus and its host cellular receptor angiotensin converting enzyme 2 (ACE2). Effective binding is dependent upon spike protein activation by transmembrane protease or furin. (B) The nasal cavity and the throat are usually the first tissues to be infected, after inhalation of viral particles. The infection of other organs where the ACE2 receptor is expressed is induced in a second stage, after the virus has colonized the upper airways. (C) Viral particles issued by replication in the nasal cavity can be transported by the air (to the lungs), in the nerves (to the brain) and possibly through the blood or lymph, by a Trojan horse mechanism (to deeper organs).

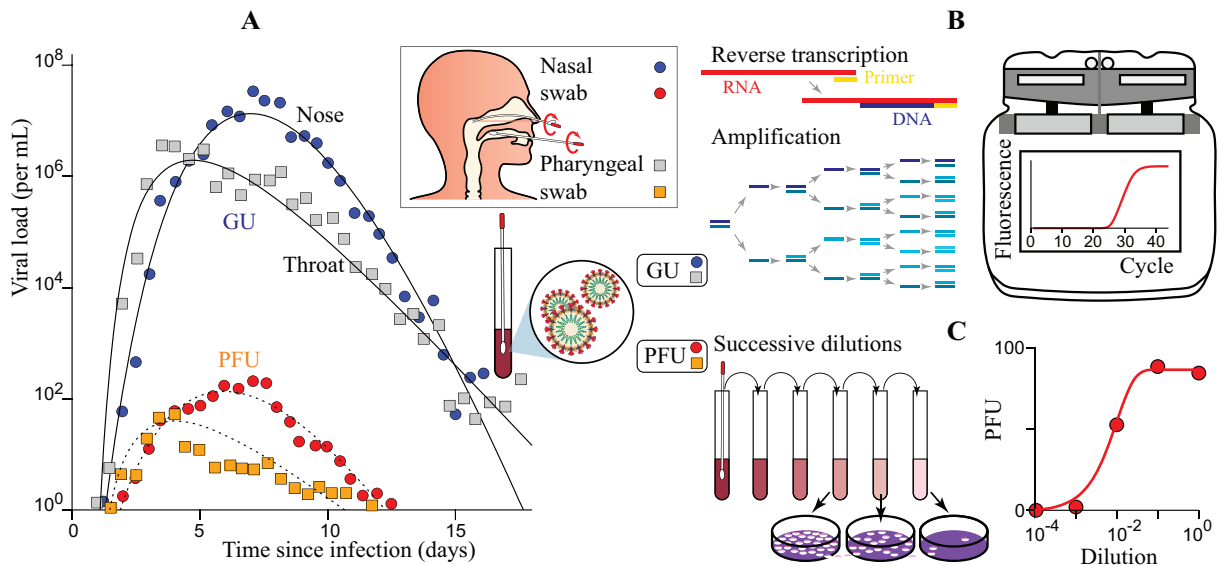

Figure 3: Viral kinetics of SARS-CoV-2. (A) Viral kinetics of SARS-CoV-2. Average viral load in the nasal cavity and the throat as a function of time since infection. Data from Killingley et al. (2022), using a strain close to Wuhan-1. (B) Viral load expressed in genome units (GU) is obtained by reverse transcription followed by quantitative PCR (RT-qPCR), which measures the quantity of viral genome copies in a viral solution. (C) Viral load expressed in plaque forming units (PFU) is measured using plaque assays, which consist counting lysis plaques in a cell monolayer in contact with a viral solution. Data of panel A are obtained for Vero Cells, using a 3 mL solution per swab.

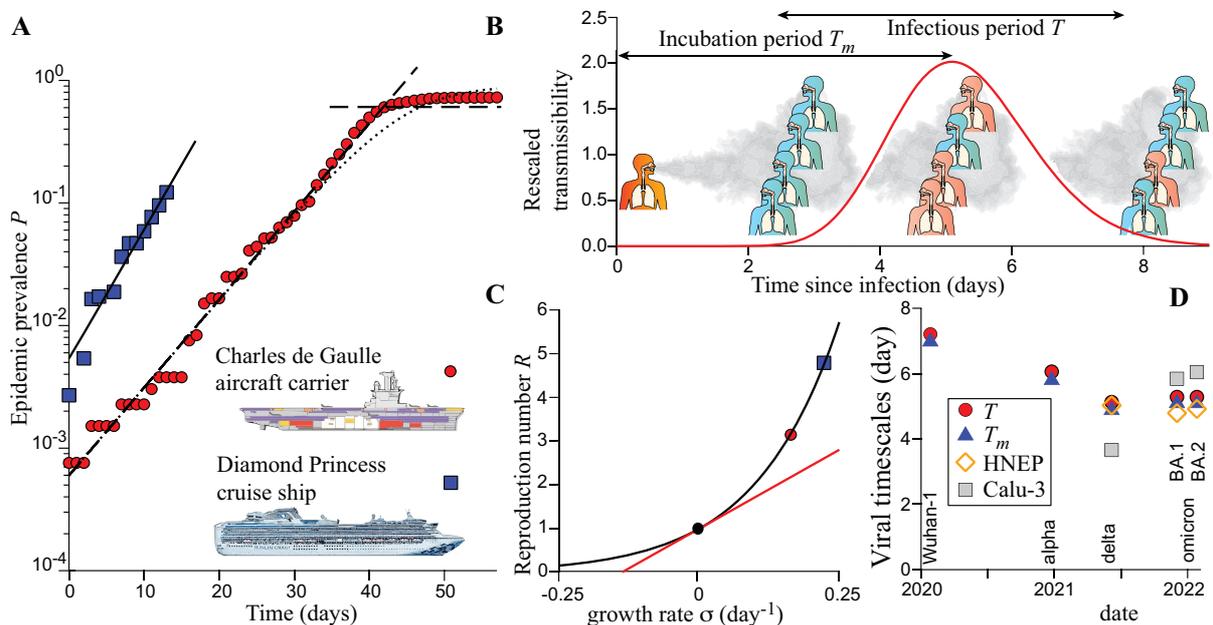

Figure 4: Calibration of the epidemic quantum. (A) Curve of the proportion $P$ of infected individuals aboard the cruise ship Diamond Princess (blue) and aboard the French aircraft carrier Charles de Gaulle (red) as a function of time $t$. The solid lines correspond to the best fit by exponential growth, resulting in a growth rate $\sigma = 0.23$ day$^{-1}$ and $\sigma = 0.17$ day$^{-1}$, respectively. The dashed horizontal line is the theoretical collective immunity limit $P \to 1/R$. The dotted line is the numerical integration of the infection equation SI-(6). On both boats, the persistence of viral particles is limited by their deactivation timescale rather than by the ventilation. The fast transmission in the Charles de Gaulle French aircraft carrier (Laval et al.,

2022) is due to the lack of filtration of the recycled air. Over 1,767 people onboard, crew and commandos together, 1,288 were infected in a short period of time. The indoor relevant volume is estimated around $150 \times 10^3$ m$^3$. The fast transmission on the Diamond Princess boat (Bazant and Bush, 2021) is due to a mostly recycling (at 70%, no HEPA filters), air conditioning, due to the cold weather conditions (−5°C). The surface accessible to passengers is $78 \times 10^3$ m$^2$ and the ceiling height is 2.4 m. The indoor volume, $187 \times 10^3$ m$^3$, is rather large compared to the number of people, N = 3,711, crew and passengers together. (B) Model dimensionless transmissibility as a function of time $t$, in days, after infection. The viral transmissibility of a person (index case) is defined as the rate of emission of replicable viral particles. In first approximation, transmissibility is proportional to the viral load in the upper respiratory tract. The incubation period $T_m$ is defined as the time between infection and maximum transmissibility, and the infectious period $T$ as the average time between infections of the index and secondary cases. (C) Relation between the epidemic reproduction rate $R$, defined as the mean number of secondary cases per index case, and the epidemic growth rate σ, predicted by the Euler-Lotka equation SI-(7), for a given transmissibility curve. It gives $R$ = 4.8 for the cruise ship and $R$ = 3.2 for the aircraft carrier. The red line is the approximation for small growth rate: $\sigma = (R - 1)/T$. (D) Incubation period $T_m$ and infectious period $T$ for the wild-type strain, Wuhan-1, and of variants Alpha (B1.1.7), Delta (B.1.617.2), Omicron BA.1 and Omicron BA.2, represented at the date of the first case in France. Viral replication timescale is measured using primary cultures of human nasal epithelial cells (hNECs), human lung cell that expresses abundant ACE2 and TMPRSS2 (Calu-3). It is defined as the time needed to multiply the viral particles by 1 billion ($10^9$).

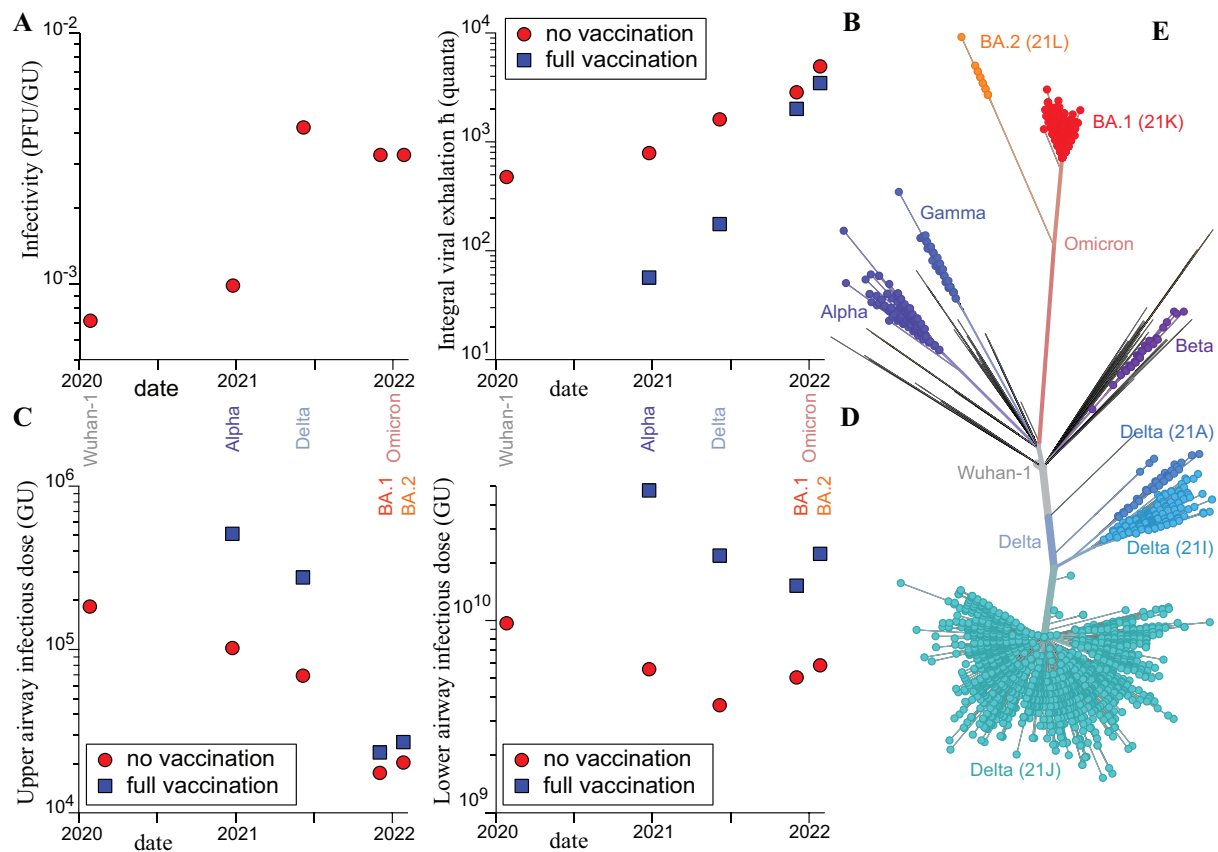

Figure 5: Epidemiologic and biological characteristics of the wild-type strain, Wuhan-1, and of variants Alpha (B1.1.7), Delta (B.1.617.2), Omicron BA.1 and Omicron BA.2. (A) Viral infectivity, defined as the average number of lysis induced by a viral particle on a Vero cell culture. It is expressed in plaque forming units per genome units (PFU/GU). (B) Mean viral

exhalation during the whole contagious time, expressed in quanta. The measurement is deduced from the variant frequency curves. (C) Infectious dose in upper airways, measured in genome units (GU). (D) Infectious dose in lower airways, measured in genome units (GU). (E) Maximum likelihood phylogenies inferred from spike nucleotide sequence; Distance corresponds to number of mutations. Source: Gisaid/ Nextstrain.

# At the crossroads of epidemiology and biology: bridging the gap between SARS-CoV-2 viral strain properties and epidemic wave characteristics.
# — Supplementary Information —

Florian Poydenot, Jacques Haiech, Alice Lebreton and B. Andreotti

January 30, 2023

**Abstract**

In this supplementary information document, we provide a self-contained review of the foundations of standard epidemiological models, aimed at being accessible with basic knowledge of physics and mathematics. The particular formulation of epidemiological equations used to construct the figures of the review has been published in a companion paper [1]. We review the assumptions and the parametrization of the model using epidemiological data. We furthermore provide the table of published measurements used in the review paper.

## 1 Epidemic growth rate and reproduction rate

Sustained airborne transmission of SARS-CoV-2 from an index case to contacts depends on biological and immune characteristics of the index case (transmissibility) and the contacts (susceptibility), but also on social and physical characteristics such as the number of available contacts, the duration of exposure, and the ventilation and mask wear during the contact (see Figure S1). We have developed the social and physical aspects of transmission in a companion paper [1], and we discuss here the calibration of biological quantities, both from a molecular and epidemiologic point of view.

Starting from the molecular description of the index case, viral kinetics determine both the transmissibility through the viral shedding rate, and the dynamics of transmission through the course of evolution of the illness. The viral kinetics is standardly described in the literature by an exponential growth of the viral load (replication and shedding) followed by an exponential decay (immunity response). However, the SARS-CoV-2 human challenge trial [2] has provided unprecedented time resolved data showing a more rounded viral load curve (Fig. 3). Here, we parametrize the viral load $V$ by the law

$$V = \mathcal{V}\psi(t) \tag{1}$$

where $\mathcal{V}$ is a characteristic concentration for a particular infected person and $\psi(t)$ the rescaled transmissibility at a time $t$ after infection (Fig. 4-B). We therefore assume that infected people with different maximum viral load present on the average the same viral kinetics up to a constant. The infectious period $T$ is defined as the average time from infection, weighted by the transmissibility:

$$T = \frac{\int_0^\infty \tau\psi(\tau)\mathrm{d}\tau}{\int_0^\infty \psi(\tau)\mathrm{d}\tau} \tag{2}$$

It is worth noting that the integrals run from $t = 0$, which is the infection time. More precisely, we choose a convenient normalization of $\psi(t)$ to ensure that it is dimensionless:

$$T = \int_0^\infty \psi(\tau)\mathrm{d}\tau \quad \text{and} \quad T^2 = \int_0^\infty \tau\psi(\tau)\mathrm{d}\tau \tag{3}$$

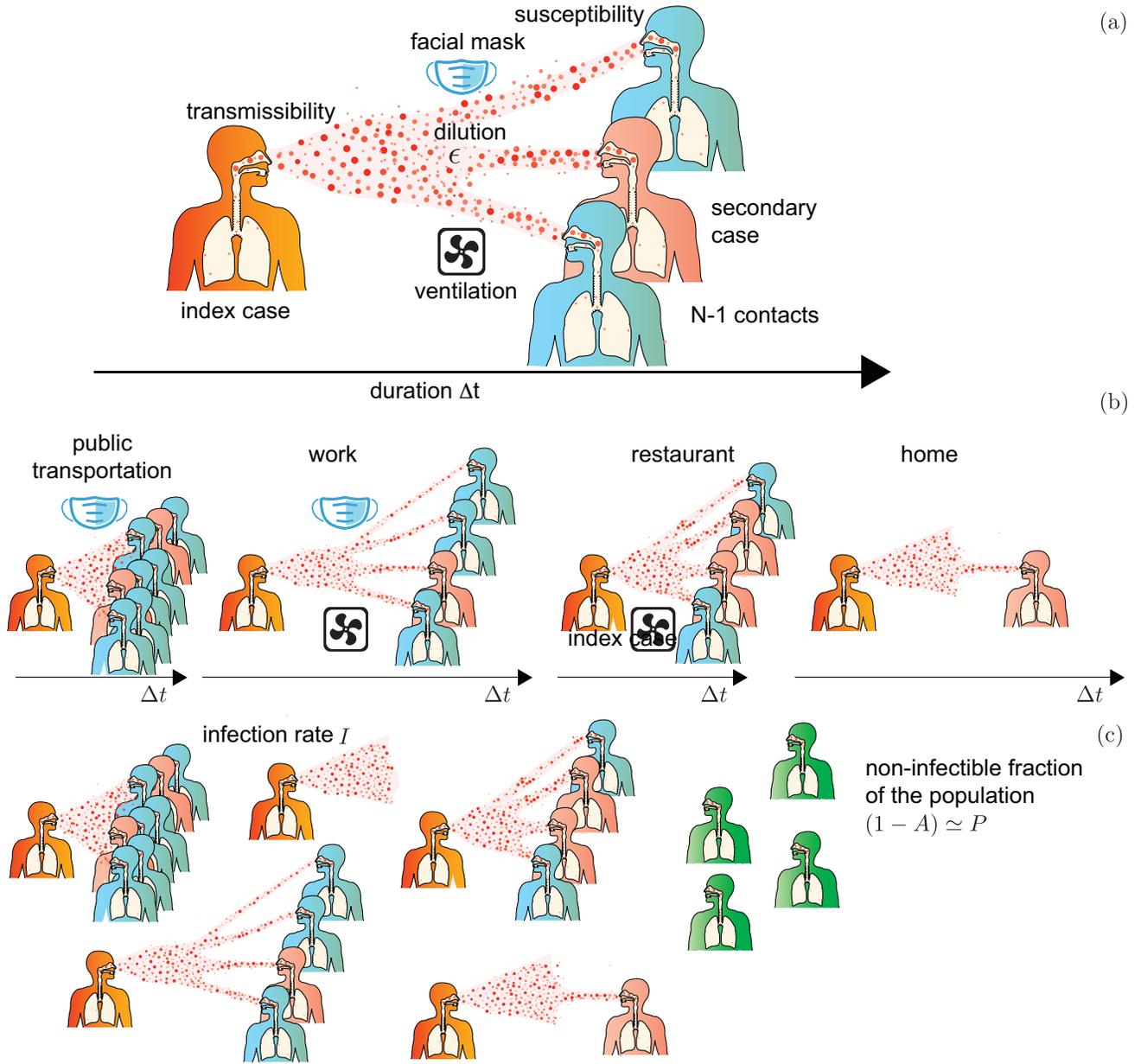

Figure S1: (a) Transmission of SARS-CoV-2 in a certain space depends on the duration $\Delta t$, on the number $N$ of people present, on the the viral emission rate (transmissibility), on the susceptibility of contacts to be infected and on the viral dilution $\epsilon$ due to ventilation and to face masks. (b) The mean number of people infected by an index case is the average over possible situations weighted by the duration $\Delta t$. Four examples are schematized. From left to right: high number $N$ but short time $\Delta t$ and moderate dilution factor $\epsilon$ (public transportation); Long time $\Delta t$ but moderate number $N$ and low dilution factor $\epsilon$ (work); Long time $\Delta t$ and high dilution factor $\epsilon$ (restaurant); Very long time $\Delta t$ with a high dilution factor $\epsilon$ but small $N$ (household transmission). (c) Transmissibility is by definition the viral emission rate, which is proportional to the replicable viral load. The infection rate $I$ is the number of new infected people per unit time. It is related to the infection rate over the past period, to the epidemic reproduction rate, and to the fraction of non-infectible people.

Here, we have introduced a phenomenological equation of the form

$$\log \psi = \log \psi_m - \frac{a(t - T_m)^2}{1 + bt} \qquad (4)$$

to fit these data with the same number of parameters. $\mathcal{V}$ is a characteristic viral load. $T_m$ is the incubation period, defined as the time from infection to the maximum viral load. The parameters $a$ and $b$ determine the initial growth rate and final decay rate. The constant $\psi_m$ is fixed by the normalization condition. We have performed a mapping of the double exponential model onto equation (4) in order to deduce the parameters that best describe the average kinetics for different variants. The resulting curves are shown in Figure S2.

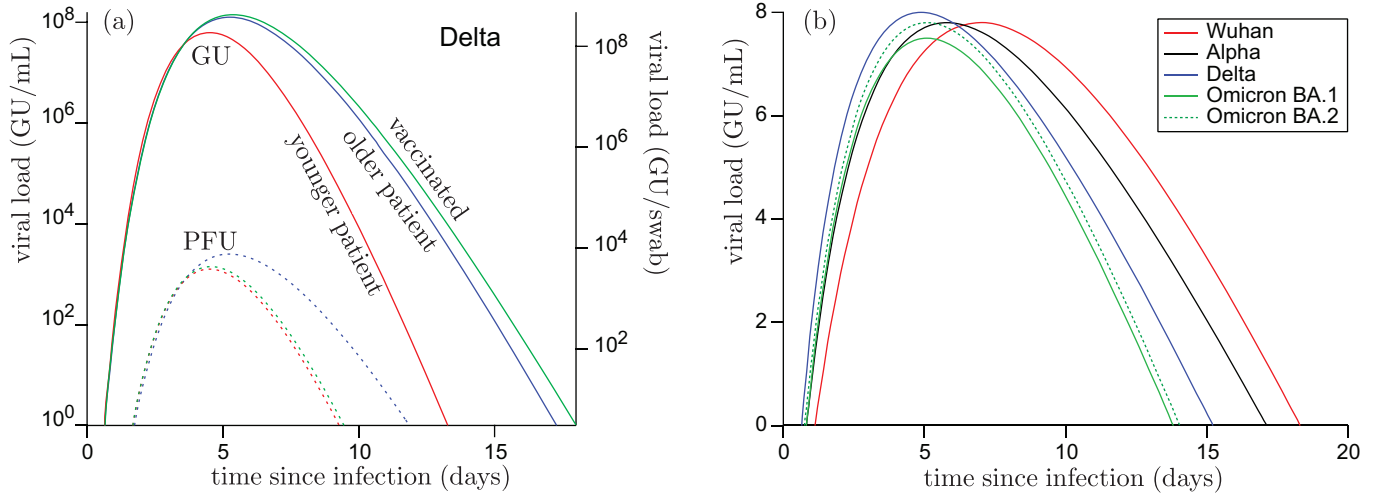

Figure S2: (a) Modelled average viral kinetics for younger and older patients, vaccinated or not, obtained by mapping the double exponential kinetics to equation (4). The characteristics are those of the Delta strain. From Néant et al. [3] (b) Modelled average viral kinetics for the successive variants of interest, obtained using the data reported in table 1. The parameters of equation (4) are reported in table 2.

The viral transmissibility of a particular infected person (the index case) is defined here as the rate of emission of replicable viral particles (Fig. S1). Under the simplifying assumption that emission is proportional to the viral load in the upper respiratory tract, the transmissibility is proportional to the viral load in the upper respiratory tract. We therefore parametrize the concentration of viral particles in the exhaled air by the law:

$$C = \mathcal{C}\psi(t) \tag{5}$$

where $\mathcal{C}$ is a characteristic concentration for a particular infected person and $\psi(t)$ the rescaled transmissibility at a time $t$ after infection (Fig. 4-B).

The infection rate $I$ is defined as the mean number of infected people per unit time in a given population. At time $t$, a secondary case infection is induced by an index case contaminated before, at a time $t - t'$, with a probability proportional to $\psi(t')$. Introducing $R$, the epidemic reproduction rate, defined as the mean number of secondary cases per index case, $I$ obeys an integral equation (see Grassly and Fraser [4] for a review and derivation):

$$I(t) = A(t)\frac{R}{T}\int_0^\infty I(t-t')\psi(t')\mathrm{d}t' \tag{6}$$

where $A(t)$ is the fraction of the population susceptible to be infected (Fig. S1). $A(t)R(t)\psi(t')/T$ is classically known as the infectiousness at time $t$ of the index case that has been infected for a duration $t'$. It is the product of the contact rate $AR/T$ at the time of contact and the biological factor of infectiousness since the onset of infection $\psi(t')$. Importantly, equation (6) remains valid even when the transmission time is comparable to the epidemic growth timescale as it takes into account the viral kinetics. $I(t)$ and $A(t)$ are unknowns, and must be solved together. When a small fraction of the population is immune, $A$ is close to 1, and the equation admits an exact exponential solution to $I(t)$, $I(t) = I_0 e^{-\sigma t}$. Plugging this into equation (6), we can relate the growth rate $\sigma$ to the epidemic reproduction rate $R$ by the Euler-Lotka equation:

$$R = \frac{\int \psi(t)\mathrm{d}t}{\int \psi(t)\exp(-\sigma t)\mathrm{d}t} \tag{7}$$

This relation is plotted in figure 4-C for the rescaled viral load shown in figure 4-B. A discussion of the Euler-Lotka renewal equation in a more general context can be found in chap. 13 of Martcheva [5].

The epidemic prevalence $P$ is defined as the fraction of the population that has been infected in the past:

$$P(t) = N^{-1} \int_{-\infty}^{t} I(t') \mathrm{d}t' \tag{8}$$

Then, assuming that each infection leads to long term immunization, a simple approximation of the susceptible fraction is $A(t) = 1 - P(t)$. However, $A$ may be very different from $1 - P$ due to vaccination, or to the gradual loss of immunity.

## 2 Reference point for the evaluation of the epidemic reproduction rate $R_0$

The epidemic reproduction rate $R$ is the average number of secondary infections per index case. It combines biological factors that determine infection susceptibility and viral transmissibility, and social factors (Fig. S1). To define a reference reproduction rate $R_0$ that would characterize the transmissibility of a given viral strain, it is necessary to choose a reference state of the social behaviors. The simplest choice is to define $R_0$ as the epidemic reproduction rate when society ignores the virus and behaves "normally". This is, by definition, only possible when the epidemic starts. Figure S3 shows the initial stage of SARS-CoV-2 epidemics in different French departments and different European countries, before the first 2020 lockdown. It can be safely assumed that the number of deaths per unit time was proportional to the number of cases, in this initial stage. The curves show the multiplicative nature of the epidemic and is direct evidence of the effect of the lockdown. The epidemic reproduction number is around the same value $R_0 = 6.9$ in both departments/countries where the epidemic arrived earlier or later. For small values of the epidemic growth rate $\sigma$, the Euler-Lotka equation can be linearized:

$$R \simeq \frac{\int \psi(t) \mathrm{d}t}{\int \psi(t)(1 - \sigma t) \mathrm{d}t} \simeq 1 + \sigma T \tag{9}$$

This popular approximation leads to a much lower value of the epidemic reproduction number, around $R_0 = 3$. This difference can be ascribed to the non-linearity in the relation between $\sigma$ and $R$, at large $\sigma T$, when the growth time $\sigma^{-1}$ is comparable or larger than the infectious period $T$.

## 3 Relation between the reproduction rate and the infectious quantum

The model described in this section is discussed in a companion paper [1] in the context of the social and physical aspects of transmission. We recall here its underlying assumptions and the calibration of the biological factors in the transmission risk. We assume here that a single viral particle initiates the infection when it penetrates a vulnerable locus where conditions are favorable. The probability that at least one viral particle manages to enter a cell and replicates is independent of the presence of others viral particles. It depends on factors such as the type of cells or the density of ACE2 receptors. Wells [6] introduced the notion of dose and quantum to describe this probability. For a person having inhaled an intake dose $d$, the probability law of infection $p(d)$ takes the form $p(d) = 1 - e^{-ad}$. $a^{-1}$ is the infection dose of the person considered. Its average over the population, $\bar{a}^{-1}$, is by definition the quantum of infection. The product $ad$ is therefore the dose, expressed in infectious quanta. At small $ad$, the probability of infection can be linearized: $p(d) \simeq ad$. This excludes super-spreading events, which occur when an infected person with a large exhaled viral concentration $C$ attends an under-ventilated place, leading to multiple simultaneous infections. The total number of exhaled viral particles is, on average, equal to $\int \bar{q} \mathcal{C} \psi(t) dt = \bar{q} \mathcal{C} T$, where $\bar{q}$ is the mean exhalation flow rate. This number can be expressed in infectious quanta to define the mean integrated quantum emission $\bar{h}$:

$$\bar{h} = \bar{q} \mathcal{C} \bar{a} T \tag{10}$$

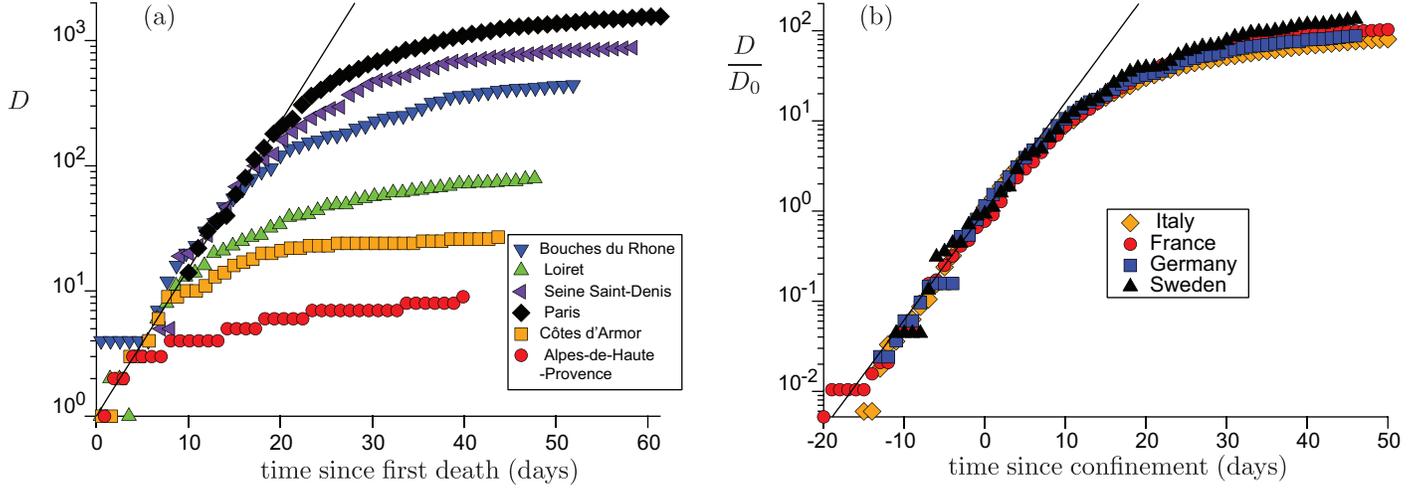

Figure S3: (a) Curve of the cumulative number of deaths $D$ as a function of time, in days, in different French departments. The time axis is shifted so as to superimpose the curves in the first phase of the epidemic on a jointly fitted exponential ($\sigma = 0.28 \text{ day}^{-1}$). For each curve, the best fit gives the time at which, statistically, the first death would have occurred on average, given the epidemic history. The further away departments are from major cities, the later the epidemic occurs and the more lockdown, imposed at the same date everywhere, has limited the number of deaths. (b) Curve of the cumulative number of COVID-19 deaths normalized by the same number, on the day of lockdown, as a function of time, in days, relative to the date of lockdown. Although the epidemic arrived at very different dates in the countries represented, the curves superimpose on a master curve, which shows the multiplicative nature of the epidemic. The best fit of the first phase of the epidemic by an exponential gives the growth rate $\sigma = 0.28 \pm 0.03 \text{ day}^{-1}$, which corresponds to $R_0 = 6.9 \pm 2$. This is much higher than the common value deduced from the linearization of the Euler-Lotka equation (9), which underestimates $R_0$ to 3.

The mean integrated quantum emission $\bar{h}$ measures the transmissibility and encodes all the biological part of the risk (Fig. S1). It is defined as an average over the sub-population attending the public space considered of the number of quanta that would have been exhaled if an infected person were there. It may depend on the particular activity taking place in the public space through the mean inhalation rate $\bar{q}$.

We consider a virtual situation in which $N$ people would stay in a certain place during their entire infectious period. Consider that an infected person amongst the $N$ people. It would exhale a dose $d = \bar{h}/\bar{a}$ or equivalently a number of infectious quanta $\bar{a}d = \bar{h}$. Introducing the dilution factor $\epsilon$ between exhalation and inhalation (Fig. S1), which characterizes the ventilation and dispersion efficiency, as well as the effect of face masks, the inhaled dose (in quanta) is $\bar{h}\epsilon$. The average secondary case number is therefore:

$$r = (N-1)\bar{a}d = (N-1)\epsilon\bar{h} \qquad (11)$$

It is proportional to the total number of exhaled infectious quanta, to the number of infectible people (Fig. S1). The epidemic reproduction rate $R$ deduces by averaging over the population, weighting the different places in which they live according to the time they spent inside (Fig. S1):

$$R = \langle \epsilon(N-1) \rangle \bar{h} \qquad (12)$$

$R$ is the product of three terms (Fig. S1): $\langle \epsilon(N-1) \rangle$ characterizes the social behaviour, including the effect of ventilation and face masks; $\bar{h}$ characterizes the biological factors. The mean integrated quantum emission $\bar{h}$ can be determined using equation (12), if social behaviours are known. Figure S4 shows the calibration of the mean integrated quantum emission using the epidemic evolution in secondary schools in the United Kingdom. We discuss the effect of masks on this determination in a companion paper [1].

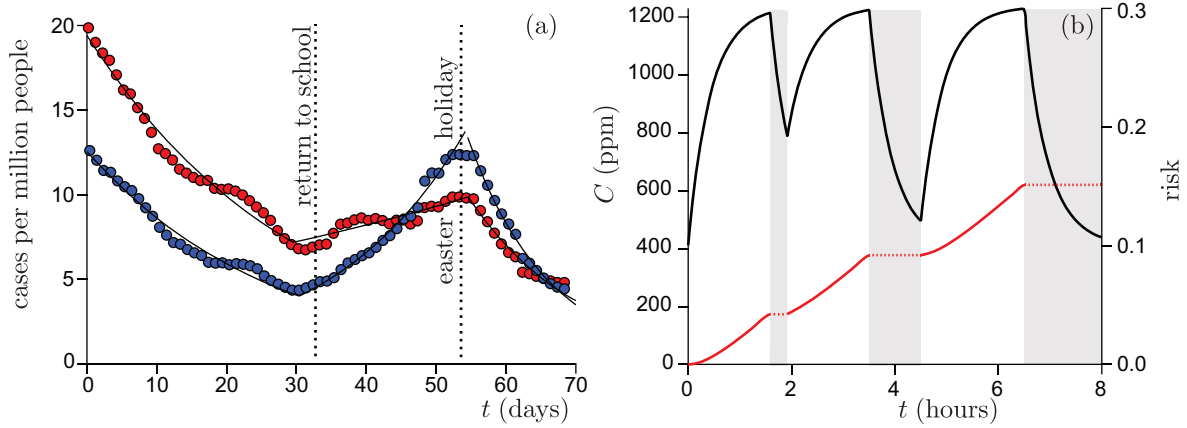

Figure S4: The epidemic evolution for secondary school pupils, between lockdown and holidays. (a) Cases per million people in the United Kingdom, from 1 February to 12 April. Blue: pupils from age 10 to age 14, with no mandatory mask. The best fit by an exponential provides the reproduction number: $R = 1.45$ ($\sigma = 0.052$ day$^{-1}$) during school period vs $R = 0.76$ before ($\sigma = -0.037$ day$^{-1}$) and $R = 0.53$ ($\sigma = -0.087$ day$^{-1}$) after. Red: pupils from age 15 to age 19, with mandatory masks. The best fit by an exponential provides the reproduction number: $R = 1.10$ during school period vs $R = 0.78$ before and $R = 0.63$ after. Uncertainties are typically 5%. The contribution of schools to the epidemic rate, in the absence of mandatory masks, is estimated to $R = 0.8$ in March 2021. (b) Typical ventilation in British schools in March, as deduced from Vouriot et al. [7], in a classroom of volume per pupil 10 m$^3$. Left axis: CO$_2$ concentration as a function of time. The pupils are not present in the classroom during the periods of time shown in gray. The average concentration is $C = 1070$ ppm. Right axis: deduced transmission risk $r$. Considering that the average school time for secondary schools is 27.5 hours per week, a total viral emission $\bar{h} = 270$ quanta is deduced. It corresponds to a typical viral emission rate $\bar{q}\bar{a}\mathcal{C} = 1.6$ quanta/hour and an emission rate at maximum on the order of 3.3 quanta/hour.

## 4 Relative transmissibility and infectivity of successive variants

Variants only interact through immunization. At low prevalence, each variant epidemic can be considered as independent from the others. The results of massive RT-PCR tests can then be used to determine the infectious quantum of successive variants of concern. Let us consider the simplest case were a variant noted $+$ replaces a variant noted $-$. The local epidemic growth rates $\sigma_-$ and $\sigma_+$ are measured during the replacement period. The infection rate of the strains are written $I_- = \mathcal{I}_- \exp(\sigma_- t)$ and $I_+ = \mathcal{I}_+ \exp(\sigma_+ t)$. The relative prevalence of the new variant therefore obeys the logistic equation:

$$\frac{I_+}{I_+ + I_-} = \frac{1}{1 + \mathcal{I}_-/\mathcal{I}_+ \exp\left((\sigma_- - \sigma_+) t\right)} \tag{13}$$

The best fit of the new variant relative prevalence by equation 13 gives the difference $\sigma_- - \sigma_+$ within a few percent uncertainty. The growth rate $\sigma_+$ of the new variant is determined by fitting the evolution of the number of new cases during the same period of time. Using the Euler-Lotka equation, the epidemic reproduction rates $R_-$ and $R_+$ are deduced. Taking the ratio $R_+/R_-$, the social component of the reproduction rate $\langle \epsilon(N-1) \rangle$ is eliminated, leaving the ratio of total viral emission $\bar{h}_+/\bar{h}_-$. For simplicity we have ignored possible small differences of immunity between the strains Wuhan-1, Alpha and Delta.

Figure S5 shows the replacement of the Wuhan strain by the Alpha strain during the winter 2021 and the replacement of the Alpha strain by the Delta strain during the summer 2021, in France. The transmissibility of the strain Alpha (resp. Delta), as measured by $\bar{h}$, is 1.7 (resp. 3.4) times larger than the wild strain. The description of the transition from the strain Delta to the strain Omicron is described in the next section.

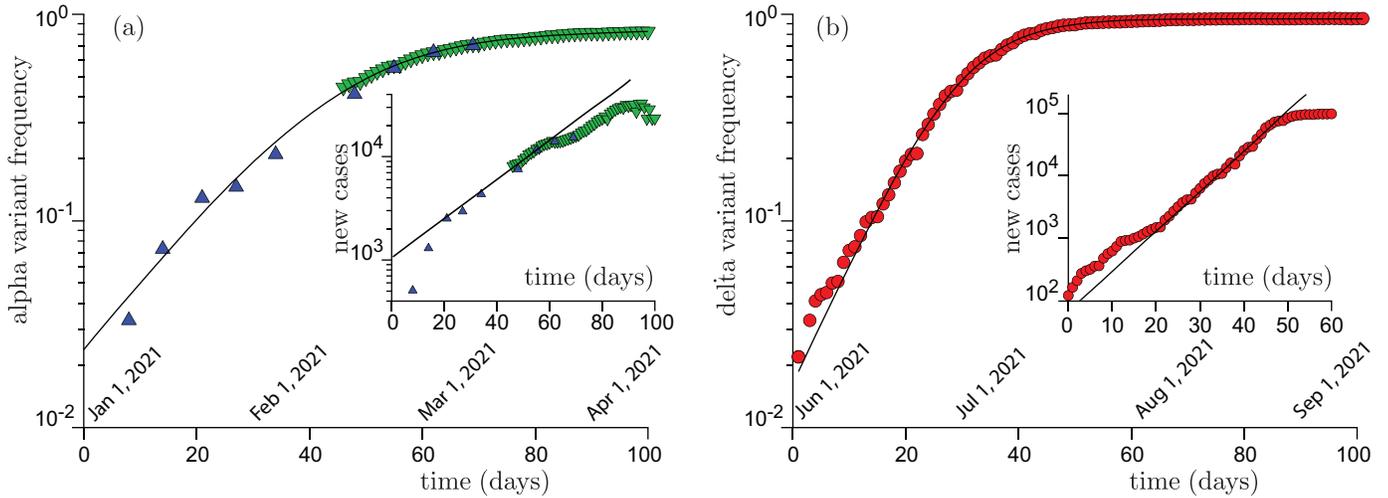

Figure S5: (a) Frequency of the variant Alpha in RT-PCR tests performed in France, as a function of time. The time origin $t = 0$ corresponds to January 1, 2021. The best fit by the logistic equation (12) gives $\sigma_+ - \sigma_- = 0.077 \text{ day}^{-1}$. Insert: number of Alpha cases identified by positive RT-PCR tests in France. The growth rate at the emergence of the alpha epidemic wave is $\sigma_+ = 0.070 \text{ day}^{-1}$. (b) Frequency of the variant Delta in RT-PCR tests performed in France, as a function of time. The time origin $t = 0$ corresponds to January 1, 2021. The best fit by the logistic equation (12) gives $\sigma_+ - \sigma_- = 0.135 \text{ day}^{-1}$. Insert: number of Alpha cases identified by positive RT-PCR tests in France. The growth rate at the emergence of the delta epidemic wave is $\sigma_+ = 0.147 \text{ day}^{-1}$.

## 5 Relative transmissibility and susceptibility of vaccinated people

The relative susceptibility $\mathcal{S}$ is by definition the ratio of the probability that vaccinated people get infected to the probability that unvaccinated, never infected people, with the same immunological characteristics, get infected. It depends on age and on the vaccination status (type of vaccine, vaccination date). The relative susceptibility can be measured from the transmission rate of sub-populations. Such measurements are well converged statistically but suffers from social biases (mask wearing, attendance of restaurants and bars, attendance of public spaces). Alternatively, it can be measured from household transmission, which removes an important bias: the vaccination status of the index case is known. On the other hand, social biases persist and the statistics is in general much lower. Other biases like age can be adjusted. However, as age, vaccination status and intrinsic susceptibility to infection (quality of the immunity) are strongly correlated, it becomes problematic to exhibit a single quantity characterizing vaccination efficiency. From the molecular biology point of view, relative susceptibility characterizes the neutralization of virus by the antigenic response. There currently exists no calibration relating molecular aspects to epidemiologic aspects.

The relative transmissibility $\mathcal{T}$ is by definition the ratio of the probability that vaccinated people with the virus (index case) infect other people (secondary cases) to the same probability for unvaccinated people. From the molecular biology point of view, the relative transmissibility can be measured as the ratio of the integral viral emission between vaccinated and unvaccinated people. From the epidemiological point of view, the relative transmissibility can only be measured through contact tracing and in particular household transmission statistics. The crude measurement is the ratio of the secondary attack rates, conditioned by the index vaccination status. The measurement suffers from social biases and a lack of statistics. Age, vaccination status and intrinsic transmissibility (quality of the immunity) are strongly correlated. It is therefore problematic to exhibit a single quantity characterizing vaccination efficiency against transmission, after an adjustment.

Figure S6 (d) shows a compilation of measurements of $\mathcal{S}$ and $\mathcal{T}$ for an up to date vaccination status. Although dispersed, the data show a clear common decrease of relative susceptibility $\mathcal{S}$ and relative trans-missibility $\mathcal{T}$. We know that they both should tend to 1 (no effect of vaccination on transmission) and to 0 (effective barrier immunity) together. A good phenomenological fit to the data is provided by the relation $\mathcal{T} = 1 - (1-\mathcal{S})^2$ and is shown in solid line in figure S6 (d).

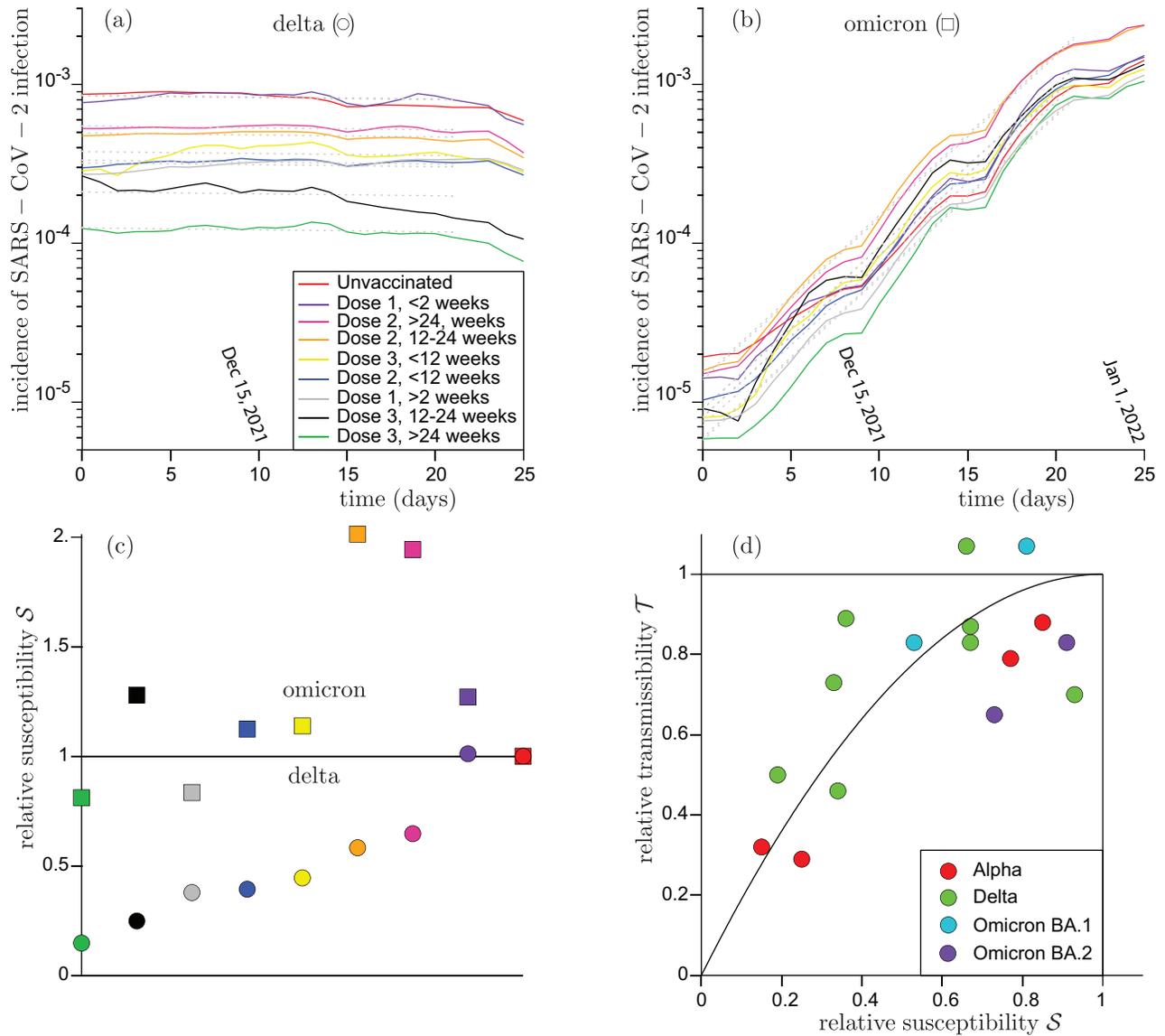

Figure S6: Incidence of SARS-CoV-2 infection for all sub-populations with different vaccination status, as a function of time, in France. (a) Raw incidence data for the Delta variant. (b) Raw incidence data for the Omicron BA.1 variant. The time origin $t=0$ corresponds to December 7th, 2021. The best fit by exponentials with the same rate for all vaccination status is superimposed. The good fit confirms that vaccinated and unvaccinated people infect each other sufficiently to share the same overall dynamics. (c) Relative susceptibility between fully vaccinated people and unvaccinated people, estimated from the relative incidence $I_j/N_j$, as a function of relative susceptibility between fully vaccinated people and unvaccinated people. (d) Relative transmissibility $\mathcal{T}$ as a function of relative susceptibility $\mathcal{S}$ for up to date vaccination status. $\mathcal{T}$ and $\mathcal{S}$ both should tend to 1 (no effect of vaccination on transmission) and to 0 (effective barrier immunity) together. Solid line: phenomenological fit $\mathcal{T} = 1 - (1-\mathcal{S})^2$.

As a simplifying assumption, we consider that contacts of vaccinated and unvaccinated people are similarly composed: then, each person is statistically in contact with vaccinated and unvaccinated people, following the fractions of the whole society. We consider a division of society into $J$ classes (according to age, vaccination status, etc.) whose size is denoted $N_j$ (such that $N = \sum_j N_j$). We denote by $\mathcal{S}_j$ and $\mathcal{T}_j$, the relative susceptibility and relative onward transmissibility associated with the class $j$. The infection rate $I_j$, defined as the mean number of infected people per unit time in the class $j$, obeys the equation:

$$I_j(t) = \mathcal{S}_j(1 - P_j(t)) \frac{R}{T} \frac{N_j}{N} \int_0^\infty \sum_k \mathcal{T}_k I_k(t - t') \psi(\tau) \mathrm{d}t' \quad \text{with} \quad T = \int_0^\infty \psi(\tau) \mathrm{d}t' \qquad (14)$$

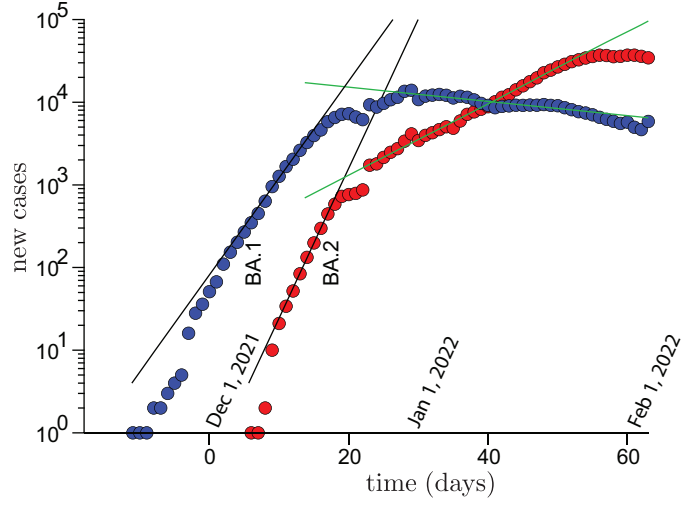

Figure S7: (a) Number of new cases of the variants BA.1 and BA.2 in France. The black and green solid lines are the best exponential fit in two different periods of time.

The epidemic reproduction rate $R$ is here defined for a virtual society of unvaccinated, never infected people. At small $P$, these equations admit an exact exponential solution of the form:

$$I_j \propto \mathcal{S}_j(1 - P_j(t)) \frac{N_j}{N} \exp(\sigma t) \qquad (15)$$

whose growth rate $\sigma$ is related to $R$ by the generalised Euler-Lotka equation:

$$R = \frac{\int \psi(t) \mathrm{d}t \, (\sum_j N_j)}{\int \psi(t) \exp(-\sigma t) \mathrm{d}t \, (\sum_j \mathcal{S}_j \mathcal{T}_j N_j)} \qquad (16)$$

Figure S6 (a) and (b) shows that incidences for all sub-populations share the same growth rate, meaning that vaccinated and unvaccinated people infect each other enough to obey the same dynamics. Under this assumption, the relative incidence $I_j/N_j$ provides an estimate of the relative susceptibility $\mathcal{S}_j$ for different vaccination schemes, displayed in Figure S6 (c). For the Delta strain, the susceptibility is ordered from $0.15$ for a complete vaccination scheme in 3 doses to $1$, the unvaccinated reference. For Omicron, on the other hand, the susceptibility is mostly $1$ within noise except for people vaccinated with two doses but delaying or refusing the third one. This may point to biases introduced by different social behaviors correlated with the vaccination status.

During three weeks, the growth rate of Omicron was $\sigma_+ = 0.23 \text{ day}^{-1}$ vs $\sigma_- = 0.0 \text{ day}^{-1}$ for Delta. The transmissibility of the strain Omicron BA.1, as measured by $\bar{h}$, is $1.8$ times larger than the Delta strain, $6$ times larger than the wild strain. Figure S7 (a) shows the number of cases of Omicron BA.1 and BA.2 in France during the replacement period. An exponential phase is observed during two short periods of time. The transmissibility of the strain Omicron BA.2 is $1.7$ larger for BA.2 than BA.1, consistently between the two estimates.

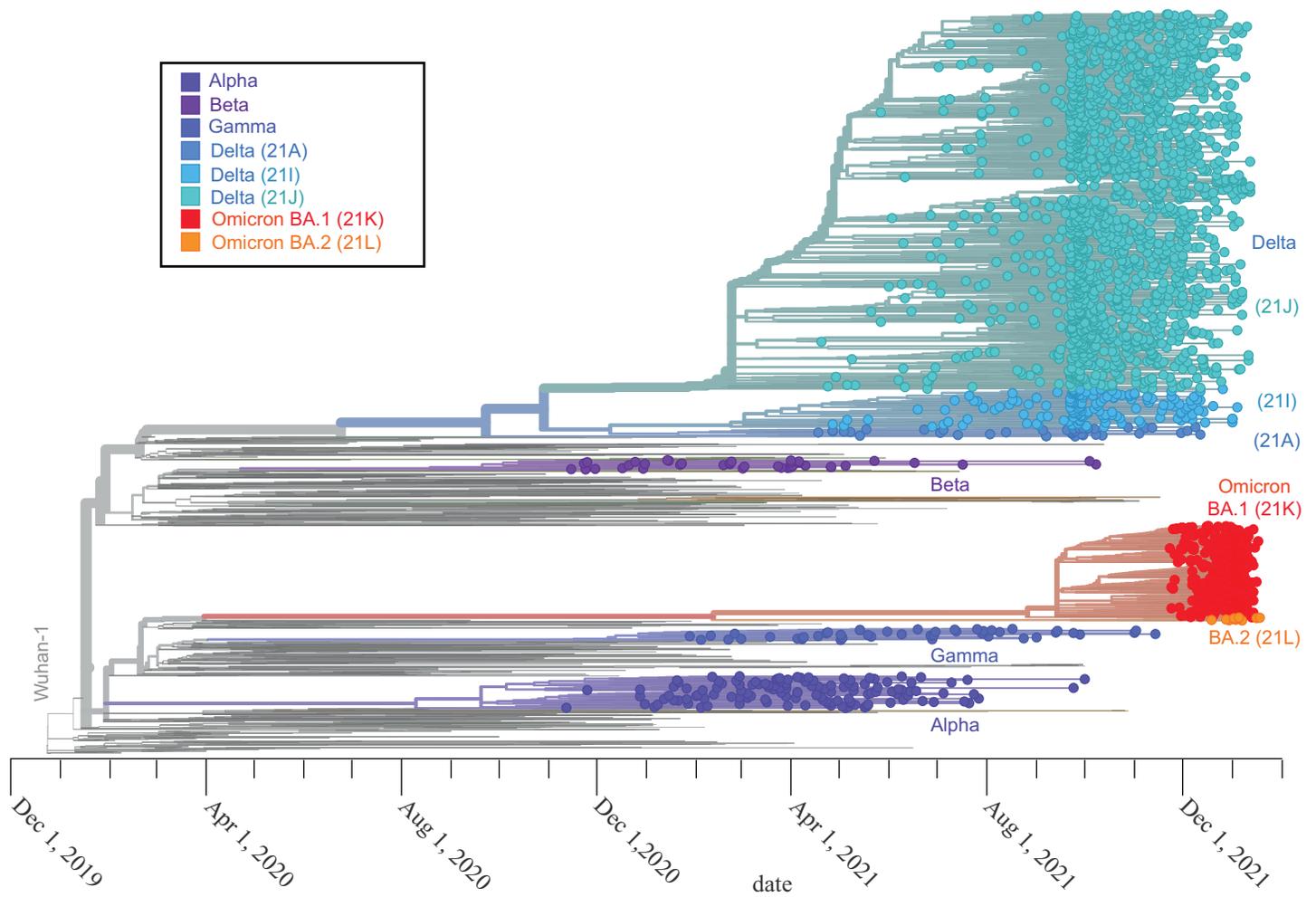

Figure S8: Phylogenetic tree up for the strains included in the article: Alpha, Beta, Gamma, Delta, Omicron BA.1 and Omicron BA.2. Source: GISAID/Nextstrain [8].

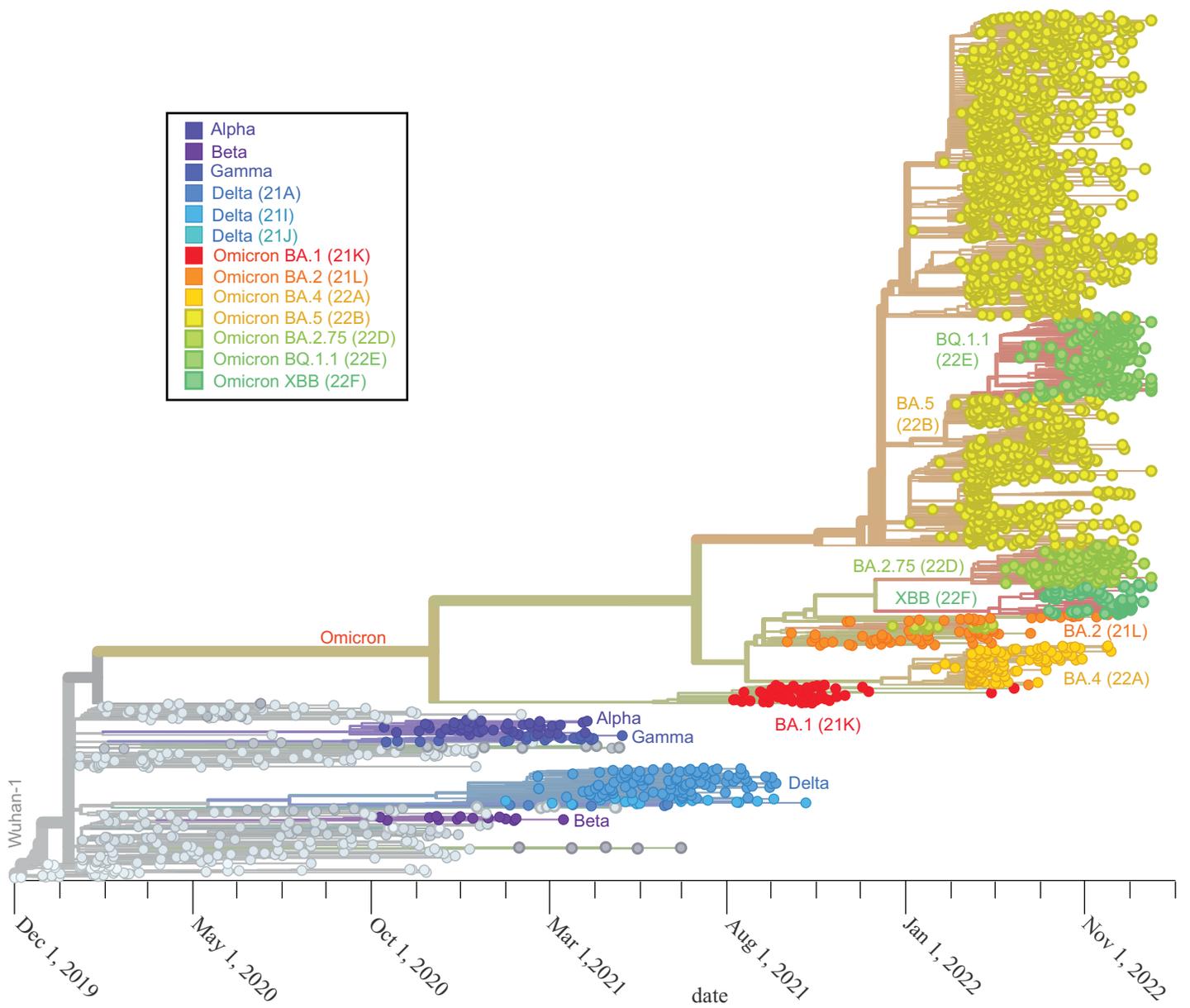

Figure S9: Phylogenetic tree including the Omicron variants: BA.1, BA.2, BA.4, BA.5, BA.2.75 and XBB. Source: GISAID/Nextstrain [8].

# 6  Supplementary Tables

Table 1: References used for viral kinetics

| Strain | Maximum viral load ($\log_{10}$ copies/mL) | Growth time (days) | Decay time (days) | Reference; comments |
|---|---|---|---|---|
| Wuhan | 7.2 | 0.35 | 0.62 | [2] |
| Wuhan | 8.1 | 0.22 | 2.7 | [9] |
| Wuhan | 8.2 | 0.30 | 0.53 | [10] |
| Wuhan | 9.7 | 0.19 | 1.2 | [3]; old people |
| Wuhan | 9.6 | 0.22 | 0.91 | [3]; young people |
| Wuhan | 7.6 | 1.4 | | [11] |
| Wuhan | 9.2 | | | [12] |
| Alpha | 8.0 | 0.26 | 0.47 | [10] |
| Alpha | 7.8 | 1.3 | | [11] |
| Delta | 7.6 | 1.1 | | [11] |
| Delta | 8.3 | 0.21 | 0.44 | [10] |
| Delta | 7.4 | | | [13] |
| Omicron | 6.9 | | | [13] |
| Omicron | 7.1 | | | [13] |
| Omicron | 7.0 | 1.2 | 0.70 | [14] |
| Unvaccinated | 8.1 | 0.26 | 0.56 | [10] |
| Vaccinated | 8.1 | 0.24 | 0.41 | [10] |
| Delta | 8.1 | 0.35 | 0.47 | [10] |
| Delta | 8.0 | | | [12] |
| Delta unvaccinated | 8.6 | | | [12] |
| Delta vaccinated | 8.2 | | | [12] |
| Delta vaccinated | 7.7 | | | [15] |
| Delta unvaccinated | 7.6 | | | [15] |
| Delta symptomatic vaccinated | 8.0 | | | [15] |
| Delta symptomatic unvaccinated | 7.9 | | | [15] |
| Delta asymptomatic vaccinated | 7.2 | | | [15] |
| Delta asymptomatic unvaccinated | 7.4 | | | [15] |
| Delta unvaccinated | 7.2 | | | [16] |
| Delta vaccinated | 7.4 | | | [16] |
| Delta vaccinated | 7.7 | | | [17] |
| Delta unvaccinated | 7.3 | | | [17] |
| Delta | 7.3 | 0.25 | 0.89 | [18] |
| Omicron BA.1 | 5.7 | 0.70 | 0.76 | [18] |

| Omicron BA.1 | 7.3 | 0.40 | 0.47 | [10] |
| Omicron BA.1 | 7.8 | | | [12] |
| Omicron BA.1 boosted | 6.9 | | | [19] |
| Omicron BA.1 unvaccinated | 7.1 | | | [19] |
| Omicron BA.2 boosted | 7.0 | | | [19] |
| Omicron BA.2 vaccinated | 7.2 | | | [19] |
| Omicron BA.2 unvaccinated | 7.4 | | | [19] |

Table 2: Parameters of the viral kinetics model

| Strain | $\log \psi_m$ ($\log_{10}$ copies/mL) | $a$ ($\log_{10}$ copies/mL) | $b$ (days$^{-1}$) | $T_m$ (days) |
| --- | --- | --- | --- | --- |
| Wuhan | 7.8 | 0.28 | 0.19 | 7.0 |
| Alpha | 7.8 | 0.40 | 0.33 | 5.8 |
| Delta | 8.0 | 0.58 | 0.40 | 4.9 |
| BA.1 | 7.5 | 0.51 | 0.30 | 5.1 |
| BA.2 | 7.8 | 0.51 | 0.30 | 5.1 |

Table 3: References used for the PFU/GU ratio

| Strain | PFU/GU | Cells | Replication | Reference |
| --- | --- | --- | --- | --- |
| Wuhan | 85000 | Vero | In vivo | [2] |
| Wuhan throat | 94000 | Vero | In vivo | [2] |
| Wuhan | 13000000 | Vero | In vivo | [12] |
| Alpha | 430000 | Vero | In vivo | [20] |
| Delta | 43000 | Vero | In vivo | [20] |
| Delta | 590000 | Vero | In vivo | [12] |
| Omicron BA.1 | 1500000 | Vero | In vivo | [12] |
| Delta unvaccinated | 1300000 | Vero | In vivo | [12] |
| Delta vaccinated | 2500000 | Vero | In vivo | [12] |
| Delta unvaccinated | 1400000 | Vero | In vivo | [12] |
| Delta vaccinated | 63000 | Vero | In vivo | [16] |
| Delta unvaccinated | 70000 | Vero | In vivo | [16] |
| Delta | 200 | hNEC | In vitro | [21] |
| Delta | 2100 | Vero | In vitro | [21] |
| Delta | 600 | Calu | In vitro | [21] |
| Omicron | 4400 | hNEC | In vitro | [21] |
| Omicron | 5600 | Vero | In vitro | [21] |
| Delta | 22000 | hNEC | In vitro | [22] |
| Omicron | 17000 | hNEC | In vitro | [22] |
| Alpha | 360 | Vero-TMPRSS2 | In vitro | [20] |
| Delta | 90 | Vero-TMPRSS2 | In vitro | [20] |

| Strain | | | | |
|---|---|---|---|---|
| Epsilon | 1400 | Vero-TMPRSS2 | In vitro | [20] |
| Alpha | 210 | Calu-3 | In vitro | [20] |
| Delta | 60 | Calu-3 | In vitro | [20] |
| Beta | $1700 - 2800$ | Vero | In vitro | [23] |
| SARS-CoV-1 Urbani | 2300 | Vero | In vitro | [24] |
| Alpha | 700 | Vero | In vitro | [25] |
| Wuhan | 6400 | Vero | In vivo | [26] |

Table 4: References used for the susceptibility and transmissibility

| Strain | Susceptibility | Transmissibility | Odds ratio type | Reference |
|---|---|---|---|---|
| Alpha | 0.15 | 0.32 | adjusted | [27] |
| Alpha | 0.34 | 0.29 | adjusted | [28] |
| Wuhan and Alpha | 0.53 | | adjusted | [29] |
| Delta | 0.15 | | raw | This study |
| Delta | 0.16 | | adjusted | [30] |
| Delta | 0.4 | | raw | [30] |
| Delta | 0.19 | 0.5 | adjusted | [27] |
| Delta | 0.34 | 0.45 | raw | [31] |
| Delta | 0.53 | 0.48 | adjusted | [32] |
| Delta | 0.61 | 1.1 | raw | [33] |
| Delta | 0.36 | 0.91 | raw | [34] |
| Delta | 0.33 | 0.71 | adjusted | [34] |
| Delta | 0.29 | | adjusted | [35] |
| BA.1 | 0.77 | | adjusted | [35] |
| BA.1 | 0.53 | | adjusted | [30] |
| BA.1 | 0.83 | | raw | [30] |
| BA.1 | 0.53 | 0.83 | adjusted | [19] |
| BA.1 | 0.67 | | raw | [19] |
| BA.1 | 0.83 | | raw | This study |
| BA.2 | 0.71 | 0.67 | adjusted | [19] |
| BA.2 | 0.77 | | raw | [19] |

Table 5: References used for the prevention of hospitalization

| Strain | Hospitalization hazard ratio | Reference; comments |
|---|---|---|
| Alpha | 1.62 | [36] |
| Alpha | 2 | [37] (from [38]) |
| Alpha | 1.7 | [39] (from [38]) |
| Alpha | 1.42 | [40] |
| Alpha | 1.52 | [41] (from [38]) |
| Alpha | 1.62 | [36] (from [38]) |
| Alpha | 1.89 | [42] (from [38]) |
| Alpha | 1.47 | [43] (from [38]) |
| Alpha | 1.6 | [44] (from [38]) |
| Alpha | 1.52 | [45] (from [38]) |
| Delta | 2.08 w.r.t. Wuhan | [41] |
| Delta | 0.97 w.r.t. Alpha | [46] |
| Delta | 2.3 w.r.t. Alpha | [47] |

| Delta | 2.8 w.r.t. Alpha | [48] |
| Delta | 1.9 w.r.t. Alpha | [49] |
| Omicron | 0.4 w.r.t. Delta | [50] |
| Omicron | 0.56 w.r.t. Delta | [51] |

Table 6: References used for the vaccine efficiency against hospitalization

| Strain | Vaccine efficiency against hospitalization | Reference |
| --- | --- | --- |
| Alpha | 4.3 | [46] |
| Delta | 3.6 | [46] |
| Omicron | 4.5 | [50] |
| Omicron | 10 | [52] |